\documentclass[11pt]{article}
\pdfoutput=1
\usepackage{color}
\usepackage{xcolor}
\usepackage{comment}
\definecolor{darkblue}{rgb}{0.1,0.1,.7}
\usepackage[colorlinks, linkcolor=darkblue, citecolor=darkblue, urlcolor=darkblue, linktocpage,hyperfootnotes=false]{hyperref} 
\usepackage{epsfig}
\usepackage{graphicx}
\usepackage{cite}
\usepackage{amsfonts}
\usepackage{amssymb}
\usepackage{bm}
\usepackage{latexsym}
\usepackage{mathtools}
	\usepackage{amsmath}
	\numberwithin{equation}{section}
	\def\bq{\begin{quote}}
		\def\eq{\end{quote}}
	 at 10truept

	\newcommand{\calo}{{\cal O}}

	\newcommand{\call}{{\cal L}}
	
	\newcommand{\calr}{{\cal R}}

\newcommand{\cR}{\mathcal{R}}

\newcommand{\cT}{\mathcal{T}}

\newcommand{\cE}{\mathcal{E}}

	\newcommand{\beq}{\begin{equation}}
		\newcommand{\eeq}{\end{equation}}
	\newcommand{\beqa}{\begin{eqnarray}}
		\newcommand{\eeqa}{\end{eqnarray}}
	\newcommand{\bea}{\begin{eqnarray}}
		\newcommand{\eea}{\end{eqnarray}}
	
	\newcommand{\hf}{\frac{1}{2}}

	\def\lesssim{~\mbox{\raisebox{-.6ex}{$\stackrel{<}{\sim}$}}~}
	\def\roughly#1{\raise.3ex\hbox{$#1$\kern-.75em\lower1ex\hbox{$\sim$}}}

\newcommand\Scatt{\Delta \cT}
\newcommand\olm{{\omega l m}}
\newcommand\ccqo{{\cal C}}
\graphicspath{{images/}}

\begin{document}

		\thispagestyle{empty}
		\begin{titlepage}
			\bigskip
			
			\bigskip\bigskip

			\bigskip
			
			\begin{center}
				{\Large \bf {Gravitational wave signatures of \\  departures from classical black hole scattering}}
				\bigskip
				\bigskip
			\end{center}

			\begin{center}

				\rm {Kwinten Fransen\footnote{\texttt{kfransen@ucsb.edu}} and Steven B. Giddings\footnote{\texttt{giddings@ucsb.edu}}}
				\bigskip \rm
				\bigskip
				
				{Department of Physics, University of California, Santa Barbara, CA 93106, USA}  \\
				\rm
				
				\bigskip \rm
				\bigskip
				
				\rm

				\bigskip
				\bigskip

			\end{center}

			\vspace{3cm}
			\begin{abstract}
		We initiate a general investigation into gravitational wave signatures of modifications to scattering of gravitational radiation from black holes.  Such modifications may be present due to the quantum dynamics that makes black holes consistent with quantum mechanics, or in other models for departures from classical black hole behavior.  
			We propose a parameterization of the corrections to scattering as a physically meaningful, model-independent, and practical bridge between
			theoretical and observational aspects of the problem; this parameterization can incorporate different models in the literature.  We then describe how these corrections influence the gravitational wave signal, {\it e.g.} of a body orbiting a much more massive black hole.  In particular, they generically change the rate of energy emission; this effect can be leveraged over many orbits of inspiral to enhance the sensitivity to small corrections, as has been noticed in simple models.  
					We provide preliminary estimates of the sensitivity of future gravitational wave observations to these corrections, and outline further work to be done to connect both to a more fundamental theory of quantum black holes, and to realistic observational situations.
									\medskip
				\noindent
			\end{abstract}
			\bigskip \bigskip \bigskip 
			
		\end{titlepage}
	
	\tableofcontents
	
	\newpage

\section{Introduction}

We have good reasons to believe that black holes in our quantum Universe behave differently from their classical idealization in general relativity.   A particularly strong reason is that a leading order attempt to incorporate quantum behavior leads to Hawing radiation\cite{Hawk}, whose description ultimately produces a massive violation of quantum-mechanical unitarity\cite{Hawkunc}.  There has been around fifty years of wide ranging debate about this problem, which  yields  the ``information paradox" or ``unitarity crisis."  There is still no consensus on the resolution to this problem, and much controversy remains.  But, one point on which there is a near consensus, represented by a wide variety of proposals,\footnote{We will not attempt a complete list of these, but a few will appear later in this paper.} is that there should be some modification to the classical physics of black holes at {\it horizon scales,} and not just at Planck distances deep within the black hole.  

We have also entered an era where we have gained observational access to the strong-field regions at the horizon scale of black holes, via two channels: very long baseline interferometry, as with the Event Horizon Telescope \cite{EventHorizonTelescope:2019dse,EventHorizonTelescope:2022wkp}, and gravitational wave detection, as with LIGO/VIRGO/KAGRA \cite{LIGOScientific:2018mvr,LIGOScientific:2020ibl,LIGOScientific:2021usb,KAGRA:2021vkt}, and in the future LISA \cite{Colpi:2024xhw}.

These two points combined suggest that we investigate possible observational signatures of new physics associated with a quantum-consistent description of black holes.  Of course, conventional wisdom holds that any new effects of quantum gravity will only appear at short, perhaps Planckian, scales, and should not be manifest in the weakly-curved vicinity of a large black hole horizon.  But, this is the same conventional wisdom that results in the inconsistency and ultimate crisis.

Different proposals and models for what could modify the black hole description of general relativity are made both at the classical level and quantum level, with greater or less motivation.  Without assessing their intrinsic merits, we can ask in what cases they could produce observational signatures.  This paper will focus on the gravitational wave case.  

Of course, if we had experimental access to a black hole, one way to probe its behavior would be to scatter radiation from it, and see whether this scattering is modified from that of the classical description; this is also the safest way to investigate a black hole.  
Instead what we have is gravitational waves from binaries involving black holes.  However, a basic point which we will develop in this paper is that gravitational wave production in a binary inspiral can be connected to gravitational wave scattering from an isolated black hole -- the binary dynamics provides a somewhat complicated source term for  gravitational waves that both scatter and contribute to the observable signal.

Specifically, we suggest that a good starting point for investigation of possible departures from classical black hole behavior is a principled parameterization of the resulting deviations in scattering behavior.  We will then investigate how to connect this to deviations in the gravitational wave signal.  Our description of scattering is model-agnostic.  We imagine that classical black holes are replaced by compact objects with a description that is consistent with quantum mechanics, ``compact quantum objects" or CQOs, to abbreviate.\footnote{We do not use the ``exotic compact object" (ECO) terminology \cite{Mark:2017dnq,Maggio:2018ivz,Cardoso:2021ehg,Ryan:2022hku,Cunha:2022gde}  both because it is commonly used to refer to horizonless modifications of black holes\cite{Cardoso:2019rvt}, and because once we understand the quantum description there should be nothing exotic about it.  In particular, it may be that CQOs deviate from classical black hole behavior only in small corrections to most physical quantities.}  Of course, the description of a CQO should have many of the features of a classical black hole, in particular since we now have various observations in which there have been no anomalous deviations from classical behavior\cite{LIGOScientific:2019fpa,LIGOScientific:2021sio,EventHorizonTelescope:2022xqj}.\footnote{There was a claim of some statistical significance for an ``echo'' observation \cite{Abedi:2016hgu} which was likely premature \cite{Ashton:2016xff,Westerweck:2017hus}.} In particular we imagine that the description of such a CQO is essentially the same as that of a classical black hole outside a radius $R_a$ which may be comparable to the horizon radius $R$, but that there are modifications to that classical behavior within $R_a$ which can alter the scattering behavior.  This  is illustrated in Fig.~1.

 \begin{figure}[t!]
 	\begin{center}
 		\includegraphics[width=0.60\textwidth]{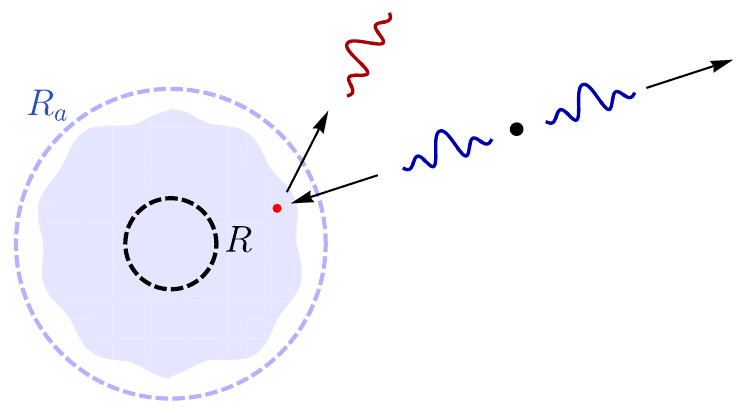} 
 		\caption{An illustration of scattering from a compact quantum object.  A source emits radiation that propagates like in a classical black hole geometry outside a radius $R_a$.  
Departures from classical black hole behavior within $R_a$ can however alter the scattering from the CQO, as compared to that of the classical geometry.  (The figure should not be taken too literally; for example
	the wavelength of the illustrated gravitational radiation is not to scale for an orbiting source; it should be significantly longer and, relatedly, the region in which it is sourced is not as localized.)}
 		\label{fig:setup}
 	\end{center}
 \end{figure} 

Propagation of classical waves in the region outside $R_a$ is then the same as that for a classical black hole, but there will be deviations in the scattering of waves from the region inside $R_a$; {\it e.g.} absorption and reflection may be modified.  For a simplified example, consider scattering of incoming massless scalar waves of definite frequency and angular momentum, which we can parameterize in the region outside $R_a$ as
\beq \label{phiscatt}
 \phi_{\omega lm}= \phi^{\rm BH}_{\omega lm} + \sum_{l'm'}\int \frac{d \omega'}{\sqrt{2 \omega'}} \Scatt_{lm,l'm'}(\omega,\omega') \phi^{up}_{\omega' l'm'}\ .
 \eeq
Here $\phi^{\rm BH}_{\omega lm}$ describes the incoming plus scattered wave from a classical BH, and we parameterize departures from this scattering via the amplitudes $\Scatt$ for corresponding purely outgoing modes $\phi^{up}_{\omega lm}$ which we will describe further later; in general the scattered waves may have different frequencies and angular momenta.  
For gravitational waves, a corresponding expression for metric perturbations is
 \beq\label{hscatt}
 H_{\omega A\mu\nu} =  H_{\omega A\mu\nu}^{\rm BH} + \sum_{A'} \int \frac{d \omega'}{\sqrt{2 \omega'}} \Scatt_{AA'}(\omega,\omega') H_{\omega' A'\mu\nu}^{up}\ ,
 \eeq
 where the indices $A,A'$ collect the other mode quantum numbers (angular momenta/polarizations).  

These expressions give a model-independent parameterization of scattering due to modified dynamics within $R_a$; in practice a similar description could be used if the object were, {\it e.g.}, a neutron star, with a correspondingly larger $R_a$.  This description extends other model independent parameterizations via multipole moments \cite{Ryan:1997hg,Barack:2006pq,Bena:2020uup,Fransen:2022jtw,Mayerson:2022ekj}, tidal deformabilities (Love numbers) \cite{Binnington:2009bb,Sennett:2017etc,Cardoso:2017cfl,Chia:2020yla}, and quasinormal modes \cite{Nollert:1999ji,Kokkotas:1999bd,Berti:2018vdi,Cardoso:2019mqo,Cano:2023jbk,Cano:2023tmv,Maselli:2023khq}.  Examples of models  for the physics altering the scattering include additional reflection near the horizon as in ``echo" models \cite{Abedi:2016hgu}\cite{Cardoso:2017cqb,Price:2017cjr,Mark:2017dnq}, other modifications to effective potentials \cite{Volkel:2017kfj,Volkel:2018czg,Volkel:2019ahb,Volkel:2019gpq}, boson clouds \cite{Baumann:2018vus,Baumann:2019ztm,Baumann:2021fkf,Baumann:2022pkl}, dirty black holes\cite{Torres:2022fyf}, and nonviolent unitarization\cite{Giddings:2011ks,Giddings:2012gc,Giddings:2013kcj,Giddings:2014nla,Giddings:2017mym,Giddings:2019vvj,Giddings:2022ipt}.  Such a principled parameterization of departures from the classical BH null hypothesis should be useful either for investigating possible deviations, or for constraining the dynamics of CQOs if deviations are found.
Of course if anomalies do present themselves, it will be a challenge to disentangle their origin, even if astrophysical sources \cite{Barausse:2014tra} and systematic modeling uncertainties can be excluded \cite{LIGOScientific:2016ebw,Purrer:2019jcp,Owen:2023mid}, and in that case such a parameterization is expected to be useful.

An effective description like this is partly motivated by a separation of scales; for example one often considers gravitational radiation with wavelength significantly longer than the horizon radius.  Moreover, details of the internal dynamics of the black hole are expected not to have large effect on coarser-grained observables such as signals from binary inspiral.   

One of our present goals is to  quantify the dependence of gravitational wave signals on any such modified dynamics, and the sensitivity to such modifications, through this generic intermediary of the modification to the scattering from a CQO.  We do so by relating the modified scattering to the gravitational wave signal from inspiral. We will describe how to do so in the extremal mass ratio case, using Green function methods; as we will discuss further, a corresponding treatment is also expected to apply perturbatively to the case of comparable masses.  This will give a relation between the generic modification to scattering parameterized by $\Scatt_{AA'}(\omega,\omega')$, and the corresponding gravitational wave signal {\it e.g.} as would be observed by LISA.  

Small deviations in scattering behavior are expected to lead to small changes in gravitational wave signals that may be hard to observe.  However, there also are possible ``amplifiers" for such deviations.  Specifically, if such a deviation alters the rate at which energy is radiated from the binary, and one considers its effect over many orbits of the inspiral, the resulting gravitational wave signal can accumulate a significant net phase shift.  This basic point was made in  \cite{Datta:2019epe}, who considered a simple model with modified reflection from a black hole; it was further investigated in \cite{Maggio:2021uge,Sago:2022bbj}.  As we will further describe, this gives an in-principle enhanced sensitivity to even relatively small departures from classical black hole scattering.  In the process we will clarify and correct the relation of energy loss to the reflection amplitude given in  \cite{Datta:2019epe}.

 Much of the basic treatment of these effects is the same for scalar radiation as for gravitational waves; scalars are important guides to understanding the basic physics.  The scalar case is simpler since one doesn't need to account to polarizations, which are typically expected to only give $\calo(1)$ corrections.  For that reason, we will give a more complete treatment in the scalar case, and leave further discussion of certain aspects of the case of gravitational radiation for future work.  The case of a nonrotating black hole is also simpler, so some of this initial exploration of the sensitivity to modifications through such dephasing also will focus on that case.

 The structure of the remainder of the paper is as follows. First, in the next section, we derive how the modifications to scattering parameterized in \eqref{phiscatt} affect a scalar wave signal arising from a source such as an orbiting body, and from that derive the change in the rate of energy emission from the orbiting source.  Following \cite{Datta:2019epe}, we parameterize this modified rate in terms of a ratio to the rate of energy absorption of a classical black hole.  Section \ref{sec:grav} then extends this analysis to the case of gravitational radiation, and derives a parameterization of the change in emitted energy in terms of an analogous ratio, as well as giving a brief discussion of the relation to a description in terms of Teukolsky variables.  Section \ref{sec:comparison} then connects our general analysis in terms of scattering amplitudes to simple examples of models for scattering.  Section \ref{sec:observational} gives a preliminary analysis of possible sensitivity of gravitational wave observations to the deviations in scattering amplitudes, through dephasing, showing significant sensitivity to small deviations.  The final section briefly discusses some of the future directions and generalizations of this work.  
Some technical aspects of the Green's functions and a time-averaging procedures are presented the appendices.

\section{Modifications to scattering, waveforms, and energy loss: scalar example}\label{sec:energyloss}

\subsection{General framework}
\label{genframe}

In an inspiralling binary, absorption and reflection of gravitational waves from the individual objects can contribute to the rate at which energy is lost from the orbital motion.  We would like to know the contribution of departures in this absorption/reflection, as described in \eqref{hscatt}, from that of classical black holes.  This is simplest to study in the case of an extremal mass ratio inspiral (EMRI), although we expect lessons to extend to the case of comparable masses.  In the EMRI limit, the smaller object can be thought of as a pointlike orbiting source for gravitational radiation, see Fig.~\ref{fig:setup}, in the background of the larger object.  To investigate the basic approach to solving this problem, we will focus on the technically simpler case of scalar radiation, \eqref{phiscatt}.

Specifically, 
consider a minimally coupled scalar field with lagrangian density
\begin{equation}\label{eqn:action}
\call = - \frac{1}{2}   \left|\nabla \Phi\right|^2 \, .
\end{equation}
We assume this describes the dynamics in a region $r>R_a>R$
where modifications to the BH geometry are assumed to be insignificant.  We initially consider the non-spinning case with $S=0$, although the discussion readily generalizes. For $r>R_a$ the geometry is then well-described as that of Schwarzschild, 
\begin{equation}\label{eqn:schwarzschild}
ds^2 = -\left(1-\frac{R}{r}\right) dt^2 + \left(1-\frac{R}{r}\right)^{-1} dr^2 + r^2 d\Omega^2_2 \, ,
\end{equation}
with $R=2M$ and $d\Omega^2_2$ the metric on the unit two-sphere. We assume that there is a source $J(x)$ in this region, which will ultimately be taken to describe coupling to the orbiting body, contributing an additional term $-J\Phi$ to the lagrangian, and resulting in the equation of motion
 \begin{equation}\label{eqn:KG}
\square\Phi =  \frac{1}{\sqrt{|g|}}\partial_{\mu} \left(g^{\mu \nu} \sqrt{|g|} \partial_{\nu} \Phi\right)	= J(x) \, ,
\end{equation} 
valid for $r>R_a$.
Our basic approach will be to solve this equation by finding a Green function, with boundary conditions at small $r$ determined by  the modification to scattering of \eqref{phiscatt}.  Then, we can calculate the total energy carried from the source by this radiation field, and its dependence on  the scattering modification $\Scatt_{lm,l'm'}(\omega,\omega')$.

Solutions to \eqref{eqn:KG} are naturally described using the partial wave expansions
\begin{equation} \label{eqn:partialwave}
\Phi(x) = \sum_{lm}\frac{u_{lm}(t,r)}{r}  Y_{lm}(\theta,\phi)\, , \quad J(x) = \sum_{lm} \frac{j_{lm}(t,r)}{r-R} Y_{lm}(\theta,\phi)\, .
\end{equation}
Then \eqref{eqn:KG} reduces to the (scalar) Regge-Wheeler equation
\begin{equation}\label{eqn:RW}
L^{\rm RW} u_{lm}\left(t,r\right) \equiv \left[\partial_{r_*}^2 - \partial_t^2 - V_l^{\rm RW}\left(r\right) \right] u_{lm}\left(t,r\right) =  j_{lm}(t,r) \, ,
\end{equation}
with 
tortoise coordinate
\begin{equation}\label{eqn:rtortoise}
\frac{d r_*}{d r} = \left(1-\frac{R}{r}\right)^{-1} \, ,
\end{equation}
and
potential 
\begin{equation}\label{RWpot}
	V_l^{\rm RW}(r) = \left(1-\frac{R}{r}\right)\left[\frac{l(l+1)}{r^2}+\frac{R}{r^3}\right] \, .
\end{equation}
We will generally suppress the angular mode indices $l$ and $m$ when there is no possibility for confusion.

The partial wave solutions of the homogeneous Regge-Wheeler equation nicely illustrate  features of the general solutions.  These depend on the boundary conditions.  Important solutions are the ``in" and ``out" solutions, characterized by their behavior at the horizon, $r_*\rightarrow -\infty$; there they take pure ingoing or outgoing forms
\begin{equation}\label{eqn:Rinoutdefsscal}
 u^{\rm in}_{\omega l}(t,r)\propto  e^{-i \omega (t+r_*)}  \quad , \quad 
 u^{\rm out}_{\omega l}(t,r)\propto e^{-i \omega (t-r_*)}  \ .
\end{equation} 
An alternative basis is that of the ``up" and ``down" solutions, which are characterized by their behavior at asymptotic infinity, $r_*\rightarrow \infty$, where they are pure outgoing or ingoing, respectively:
\begin{equation}\label{eqn:Rupdowndefsscal}
u^{\rm up}_{\omega l}(t,r)\rightarrow e^{-i \omega (t-r_*)} \quad , \quad 	u^{\rm down}_{\omega l}(t,r) \rightarrow  e^{-i \omega (t+r_*)} .
\end{equation} 

\begin{figure}[t!]
	\begin{center}
		\includegraphics[width=0.48\textwidth]{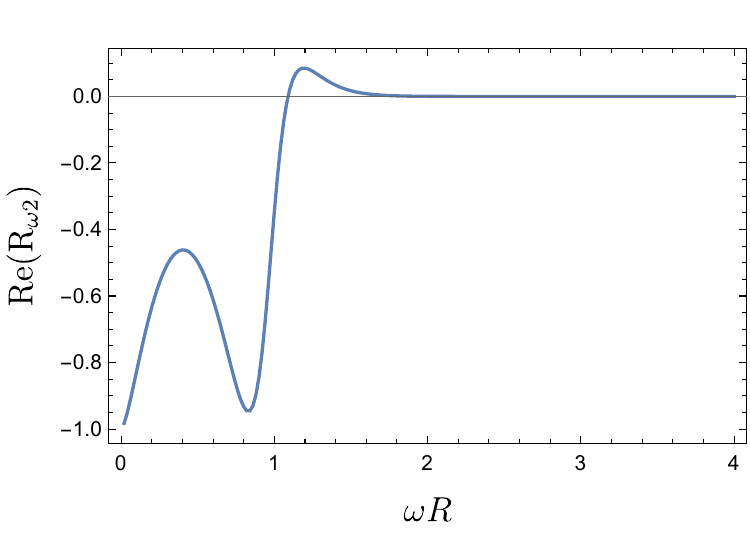} 
		\includegraphics[width=0.48\textwidth]{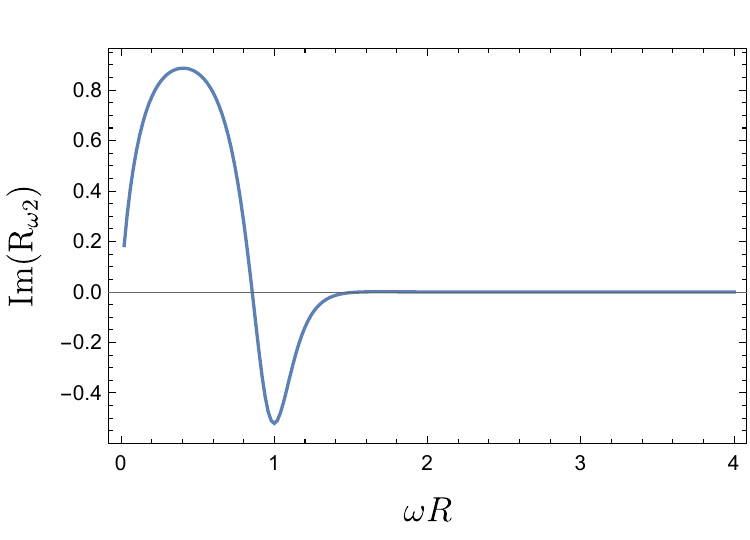} 
		\caption{Real (left) and imaginary (right) parts of the reflection amplitude for a black hole as defined in \eqref{ininfty} for the mode $l=2$.}
		\label{fig:BHRandT}
	\end{center}
\end{figure}

The in modes are appropriate for describing scattering where an incident unit-amplitude wave scatters from a classical BH, and with this normalization take the form 
\beq\label{ininfty}
u^{\rm in}_{\omega l}(t,r)\rightarrow e^{-i \omega (t+r_*)} + R_{\omega l} e^{-i \omega (t-r_*)} \, ,
\eeq
in the asymptotic region $r_*\rightarrow \infty$, with $R_{\omega l}$ a reflection coefficient due to the potential \eqref{RWpot}; the out modes are likewise chosen to have a unit-amplitude up wave in this region.  Then, approaching the horizon, $r_*\rightarrow -\infty$,
\beq\label{intohor}
u^{\rm in}_{\omega l}(t,r)\rightarrow T_{\omega l}e^{-i \omega (t+r_*)}\ ,
\eeq
where $T_{\omega l}$ is the transmission coefficient through the potential barrier.
With our normalization convention, the up modes behave at the horizon, $r_*\rightarrow-\infty$, as 
\beq
u^{\rm up}_{\omega l}(t,r)\rightarrow \frac{ e^{-i \omega (t-r_*)} + \tilde R_{\omega l} e^{-i \omega (t+r_*)}}{T_{\omega l}} \, ,
\eeq
where we introduce an internal reflection coefficient $ \tilde R_{\omega l}$.
The relation between the sets of modes is then given by the transmission and reflection coefficients for the potential barrier surrounding the BH.  For example, with the preceding normalizations
\beq\label{indownreln}
u^{\rm in}_{\omega l}(t,r) = u^{\rm down}_{\omega l}(t,r) + R_{\omega l} u^{\rm up}_{\omega l}(t,r)\, .
\eeq
In Figure \ref{fig:BHRandT} we illustrate the functions $R_{\omega l}$ for the mode $l=2$.
Note also that with the definition $u_{\omega l}(t,r)=\exp\{-i\omega t\} u_{\omega l}(r)$, the spatial wavefunctions satisfy
\beq\label{ccreln}
u_{\omega l}^{in*}(r) = u_{\omega l}^{out}(r)\quad ;\quad  u_{\omega l}^{up*}(r) = u_{\omega l}^{down}(r)\ .
\eeq

Next, if we consider modifications to this scattering due to CQO corrections to the classical geometry, we can parameterize their form in the region $r>R_a$ where the corrections are negligible in terms of modifications to the reflection/transmission, as was described in the introduction:
\beq\label{scattbc}
\phi^{sc}_{\omega lm}(x) = \frac{u_{\omega l}^{\rm in}(t,r)}{r}Y_{lm}e^{-i\omega t} + \sum_{l'm'}\int_0^\infty \frac{d \omega'}{\sqrt{2 \omega'}} \Scatt_{lm,l'm'}(\omega,\omega')   \frac{u_{\omega' l'}^{\rm up}(t,r)}{r}Y_{l'm'}e^{-i\omega' t}  \ ,
\eeq
where we include the possibility of a scattered contribution with unequal frequency and $l,m$.
A special case is the ``elastic" case, with equal frequencies, which we parameterize as
\beq\label{efscatt}
u^{sc}_{\omega l} = u_{\omega l}^{\rm in} +\Scatt_l(\omega)   u_{\omega l}^{\rm up}\ .
\eeq

To determine how the scattering correction $\Scatt$ influences the waveform and the rate at which a body orbiting the CQO at $r>R_a$ radiates energy in $\phi$, we first find a Green function for the problem \eqref{eqn:KG} with the modified boundary condition \eqref{scattbc}.  If this Green function satisfies the equation
\beq\label{GFeq}
\square G(x,x')= \frac{\delta^4(x-x')}{\sqrt{|g|}}\ ,
\eeq
the solution to \eqref{eqn:KG} is
\beq\label{PhiJ}
\Phi_J(x)= \int dV_4' G(x,x') J(x')\ ,
\eeq
with $dV_4=\sqrt{|g|} d^4x$.  

Let $G^{bh}(x,x')$ denote the Green function with the BH boundary conditions, {\it i.e.} $\Scatt=0$; for $r<r'$ this can be expanded in the form
\beq\label{BHgreen}
G^{bh}(x,x') = \sum_{lm} \int_0^\infty d\omega K^<_{\omega l m}(x') \phi^{in}_{\omega lm}(x) + cc\ ,
\eeq
where we use the general definition 
\beq\label{phidef}
\phi_{\omega lm}(x) =e^{-i\omega t}  \frac{u_{\omega l}(r)}{r} Y_{lm}(\theta, \phi) \, ,
\eeq
and {\it cc} denotes complex conjugate.
Using a matching procedure to  solve for the BH Green function (see appendix \ref{app:BHGF}) gives the expression
\beq\label{bhkl}
K^<_{\omega lm}(x') = \frac{1}{4\pi i\omega} \phi^{down*}_{\omega l m}(x') ,
\eeq
and the $r>r'$ expression
\beq\label{GFout}
G^{bh}(x,x') =\sum_{lm} \int_0^\infty d\omega \frac{\phi^{out*}_{\omega lm}(x') \phi^{up}_{\omega l m}(x)}{4\pi i\omega} + cc\ .
\eeq

The difference between any two Green functions is a solution of the homogeneous equation.  Given the BH result \eqref{BHgreen}, \eqref{bhkl}, the expression
\beq\label{PerturbedG}
G^{sc}(x,x')= G^{bh}(x,x') + \left[\sum_{ll'mm'} \int_0^\infty \frac{d\omega d\omega'}{\sqrt{2\omega'}} \frac{\phi^{down*}_{\omega lm}(x')}{4\pi i\omega} \Scatt_{lm,l'm'}(\omega,\omega')\phi^{up}_{\omega' l'm'}(x) + cc\right]\ ,
\eeq
valid for both $r<r'$ and $r>r'$, satisfies both the Green function equation  \eqref{GFeq} and the boundary conditions \eqref{scattbc}, and so is the correct Green function with the CQO boundary conditions; to reduce notational clutter in subsequent formulas we restrict to the equal $l$ case, which then easily generalizes.
This in turn gives the solution
\beq\label{newsoln}
\Phi_J(x)= \Phi_J^{bh}(x) +  \left\{\sum_{lm}\int_0^\infty \frac{d\omega d\omega'}{\sqrt{2\omega'}} Z^{down}_{\omega lm}[J]  \Scatt_l(\omega,\omega') \phi^{up}_{\omega' lm}(x) + cc\right\} \equiv \Phi_J^{bh}(x) + \Delta\Phi_J(x) \, ,
\eeq
to the sourced equation \eqref{eqn:KG}, where we define
\beq\label{Zdefs}
Z^{down}_{\omega lm}[J]= \int dV_4 \frac{\phi^{down*}_{\omega lm}(x)}{4\pi i\omega} J(x)\quad {\rm and} \quad Z^{up}_{\omega lm}[J]= \int dV_4 \frac{\phi^{up*}_{\omega lm}(x)}{4\pi  i\omega} J(x) .
\eeq
The  black hole contribution as $r \to \infty$  is given by
\begin{equation}\label{BHcont}
\Phi_J^{bh}(x) \to \sum_{lm}\int d \omega \, Z^{out}_{\omega lm}[J] \phi^{up}_{\omega l m}(x) + cc \, , \quad Z^{out}_{\omega lm}[J]= \int dV_4 \frac{\phi^{out*}_{\omega lm}(x)}{4\pi i\omega} J(x) .
\end{equation}
The coefficients $Z^{out}_{\omega lm}[J]$, $Z^{up}_{\omega lm}[J]$, $Z^{down}_{\omega lm}[J]$ are not independent; writing the defining integrals in terms of $t \to -t$, $\phi \to -\phi$, and using \eqref{ccreln} and \eqref{indownreln}, gives
\beq\label{eqn:Zrel}
Z^{up}_{\omega lm}[J]= Z^{out}_{\omega lm}[J] - Z^{down}_{\omega lm}[J] R_{\omega l}\ .
\eeq

Eq.~\eqref{newsoln} describes how to find the change in the scalar waveform for a given source, {\it e.g.} corresponding to an orbiting body, in terms of the modifications to scattering parameterized by $\Scatt_l(\omega,\omega')$. One important effect of this change  is a change in the rate of energy emission; as we will discuss, this can accumulate to enhance sensitivity to small scattering corrections.

To find how the boundary condition of \eqref{scattbc} modifies the energy emitted by the source, we compute the stress tensor 
\beq\label{stressT}
T_{\mu\nu}=\nabla_\mu\Phi \nabla_\nu\Phi - \hf g_{\mu\nu} \nabla^\lambda\Phi\nabla_\lambda \Phi \, ,
\eeq
of the scalar radiation.  For a free theory, this is a bilinear expression in the field, which with \eqref{newsoln} becomes
\beq
T_{\mu\nu}(\Phi_J,\Phi_J) = T_{\mu\nu}(\Phi_J^{bh},\Phi_J^{bh}) + T_{\mu\nu}(\Delta\Phi_J,\Delta\Phi_J) + 2 T_{\mu\nu}(\Phi_J^{bh},\Delta \Phi_J)\ ,
\eeq
where we have used symmetry of the bilinear.  
Let the source be bounded by  inner (outer) radii $r_1$ ($r_2$); the net power emitted from this region containing the source $J$ is then given by
\beq
-\frac{dE}{dt} = -\left(\int_{r_2} r^2 d\Omega - \int_{r_1} r^2 d\Omega\right) T^r_{t}(\Phi_J,\Phi_J) \ .
\eeq
However, since $\Delta \Phi_J$ is a homogeneous solution, 
\beq
\nabla^\mu T_{\mu\nu}(\Delta\Phi_J,\Delta\Phi_J)=0\ ,
\eeq
and so this term has no net effect on the emitted power; in steady state  the two flux contributions cancel.  
As a result the change in the power emitted due to the CQO boundary conditions is linear in $\Delta \Phi_J$, and hence in $\Scatt_l$, and is given by
\beq\label{Empower}
-\Delta \frac{dE}{dt} = -2\left(\int_{r_2} r^2 d\Omega - \int_{r_1} r^2 d\Omega\right) T^r_{t}(\Phi_J^{bh},\Delta \Phi_J)\ .
\eeq
The linearity arises from the change in power being an interference effect.

From \eqref{BHgreen} and \eqref{indownreln}, we see that for $r<r_1$, the black hole solution $\Phi_J^{bh}$ has a direct down piece
\beq
\phi_J^{bh, <direct}= \sum_{lm} \int_0^\infty d\omega Z_{\omega lm}^{down}[J] \phi^{down}_{\omega lm}(x) +cc\, , 
\eeq
and a reflected up piece
\beq\label{refup}
\phi_J^{bh, reflect}= \sum_{lm} \int_0^\infty d\omega Z_{\omega lm}^{down}[J] R_{\omega l} \phi^{up}_{\omega lm}(x) +cc\ .
\eeq
And, for $r>r_2$, we see that $\Phi_J^{bh}$ has a direct up piece, 
\beq\label{phidirJ}
\phi_J^{bh, >direct}= \sum_{lm} \int_0^\infty d\omega Z_{\omega lm}^{up}[J] \phi^{up}_{\omega lm}(x) +cc\ .
\eeq
plus the continuation of the reflected up piece, \eqref{refup}.  Consider the contributions of these different pieces to \eqref{Empower}.

First consider the expression $T_{rt}(\Phi_J^{bh, direct},\Delta \Phi_J)$ at $r_1$.  If we expand in terms of contributions of definite frequency,
\bea
\phi_{\omega_1}&=& \phi_1(r) e^{-i \omega_1 t} + \phi_1^*(r) e^{i \omega_1 t}\cr 
\phi_{\omega_2}&=& \phi_2(r) e^{-i \omega_2 t} + \phi_2^*(r) e^{i \omega_2 t}\ ,
\eea
then
\bea
T_{rt}(\phi_{\omega_1},\phi_{\omega_2})&=& \hf\left(\partial_r \phi_{\omega_1} \partial_t \phi_{\omega_2} + \partial_r \phi_{\omega_2} \partial_t \phi_{\omega_1}\right)\cr
&=& \frac{i}{2} \left( \omega_2 \partial_r \phi_1\phi_2^* - \omega_1 \partial_r\phi^*_2\phi_1\right) e^{-i(\omega_1-\omega_2) t} + (\cdots)e^{-i(\omega_1+\omega_2)t} + cc\ .
\eea
We will focus on the time-averaged power, over a long time $T$.  With this averaging, the second expression on the right vanishes.  The first term has average given in terms of
\beq\label{Deltadef}
\Delta_T(\omega_1-\omega_2) = \frac{1}{T} \int_{-T/2}^{T/2} dt e^{-i(\omega_1 -\omega_2) t}\ ;
\eeq
in the long time limit, this is $\propto \delta(\omega_1-\omega_2)$.  (For further discussion  of time averaging, see Appendix \ref{app:averaging}.)
In this approximation where the frequencies are equal,
\beq\label{Twrons}
\langle T_{rt}(\phi_{\omega_1},\phi_{\omega_2})\rangle \simeq -\frac{i\omega_1}{2}W_r(\phi_1, \phi_2^*) \Delta_T(\omega_1-\omega_2)+ cc\ ,
\eeq
where $W_r$ is the Wronskian, defined by
\beq
W_r(\phi_1,\phi_2)= \phi_1\partial_r\phi_2 -\partial_r\phi_1 \phi_2 = \frac{W_r(u_1,u_2)}{r^2}\ ,
\eeq
with $\phi_i = u_i/r$ as in \eqref{eqn:partialwave}.
The angular integral likewise equates the angular quantum numbers.  Thus this contribution only has terms  proportional to
\beq
W_r(u_{\omega lm}^{down}, u_{\omega lm}^{up*})\ ,
\eeq
which vanishes as can be seen by constancy of the Wronskian $W_{r_*}= W_r(dr/dr_*)$
and its evaluation at $r_*=\infty$ using the boundary conditions \eqref{eqn:Rupdowndefsscal}.

Next consider the expression $T_{rt}(\Phi_J^{bh, reflected},\Delta \Phi_J)$.  We again consider the case of definite frequency components;  these frequencies are again equal for a nonzero contribution to the time average, and angular integration equates the angular momenta.  So, we again find \eqref{Twrons}, with both arguments now being up solutions.  This expression is conserved by the equations of motion, and so for it the two terms of \eqref{Empower} cancel, as can also be explicitly seen from the behavior of the Wronskian.

The result is that the change in average radiated power is 
\beq
-\Delta\Big\langle\frac{dE}{dt}\Big\rangle = -2 \int_{r_2} r^2 d\Omega \langle T^r_{t}(\phi_J^{bh, >direct}, \Delta \Phi_J)\rangle\ .
\eeq
To evaluate this, combine \eqref{phidirJ} with 
\eqref{newsoln}, which then gives nonvanishing contribution 
\beq\label{DeltaPav}
-\Delta\Big\langle\frac{dE}{dt}\Big\rangle =- \sum_{lm} \int_0^\infty d\omega d\omega' \frac{d\omega''}{\sqrt{2\omega''}} Z_{\omega lm}^{up *}[J] Z^{down}_{\omega' lm}[J] \Scatt_{l}(\omega',\omega'') \langle  2r^2 T^r_{t}(\frac{u^{up *}_{\omega l}}{r}, \frac{u^{up }_{\omega'' l}}{r})\rangle_{\vert_{r_2}} + cc\ .
\eeq
Then \eqref{Twrons}, together with evaluating the constant Wronskian $W_{r^*}$ at infinity,
\beq
W_{r*}(u^{up*}_{\omega l} ,u^{up}_{\omega l} ) = 2 i\omega\ ,
\eeq
give the result 
\beq\label{DeltaP}
-\Delta\Big\langle\frac{dE}{dt}\Big\rangle =\sum_{lm} \int_0^\infty d\omega d\omega' \frac{d\omega''}{\sqrt{2\omega''}} 4\omega^2  Re\left\{ \Delta_T(\omega-\omega'')  Z_{\omega lm}^{up *}[J] Z^{down}_{\omega' lm}[J]\, \Scatt_{l}(\omega',\omega'')\right\} \ .
\eeq

The formula \eqref{DeltaP} is a general expression giving the change to the emitted power in terms of the quantities $\Scatt_l(\omega,\omega')$ parameterizing CQO modifications to classical black hole scattering.  While it has been derived in the example of Schwarzschild, the same derivation applies for black hole spin  $S>0$, with an analogous result.  For long-time averages, $\Delta_T(\omega-\omega')$ becomes a delta function on frequencies (see  Appendix \ref{app:averaging} for further discussion of time averaging), and the factors $Z^{up}[J]$ and $Z^{down}[J]$ are overlaps of the source $J$ with the respective wavefunctions for the classical BH, given in \eqref{Zdefs}.  As we have noted, the change of radiated power is linear in $\Scatt$, one explanation being this is due to an interference effect.  As we will discuss later, this is important for potential observability of the modifications, since for a small modification $\Scatt$ a quadratic effect would be more highly suppressed.\footnote{This differs from an assumption relating the change in power to the square of the change of absorption.  Such a parameterization was effectively taken in  \cite{Datta:2019epe}, although in the end this only affects the interpretation of their results.}

We can decompose $\Delta \cal T$ into equal and unequal frequency pieces,
\beq\label{Tdecomp}
\Scatt_l(\omega',\omega) = \sqrt{2\omega} \Scatt_l(\omega) \delta(\omega-\omega') + \Scatt^{\neq}_l(\omega',\omega)\ ,
\eeq
where 
\begin{equation}
	\lim_{\epsilon \to 0} \int^{\omega+\epsilon}_{\omega-\epsilon} d\omega' \,  \Scatt_l^{\neq}(\omega,\omega') = 0\, .  
\end{equation}
In the pure equal-frequency case, \eqref{DeltaP} simplifies to
\beq\label{DeltaPeq}
-\Delta\Big\langle\frac{dE}{dt}\Big\rangle = \sum_{lm} \int_0^\infty d\omega d\omega' 4 \omega^2  Re\left\{ \Delta_T(\omega-\omega')  Z_{\omega lm}^{up *}[J] Z^{down}_{\omega' lm}[J]\, \Scatt_{l}(\omega')\right\} \ .
\eeq

A useful phenomenological parameter is the ratio of the 
change in the energy loss given by either \eqref{DeltaP} or \eqref{DeltaPeq} to  the total rate of energy loss in the BH case, \eqref{EBHtot},
\beq\label{rattot}
\Delta\Big\langle\frac{dE}{dt}\Big\rangle\Big/ \Big\langle\frac{d E}{d t}\Big\rangle_{BH,tot} \, ,
\eeq
 or, as in \cite{Datta:2019epe},  its ratio to the energy loss \eqref{EBHhor} to the would-be BH horizon,
\beq\label{ratbh}
\ccqo=\Delta\Big\langle\frac{dE}{dt}\Big\rangle\Big/ \Big\langle\frac{d E}{d t}\Big\rangle_{BH,hor} \, .
\eeq

Time averaging for more general sources is discussed in Appendix \ref{app:averaging}, but consider the special case where the source can be approximated as having discrete frequencies, as with a periodic orbit:
\beq
J=\sum_{n=0}^{\infty} J_n(\vec x) e^{-i\omega_n t}\ + cc\ .
\eeq
In this case
\beq
Z_\olm[J] = \sum_n Z_{nlm}[J] \delta(\omega-\omega_n)\ ,
\eeq
where 
\beq Z_{nlm}[J] = \int \sqrt{|g_{tt}|}dV_3 \frac{u^*_{\omega_n l}(r)}{2i\omega_n r} Y^*_{lm}(\theta,\phi) J_n(\vec x)\ .
\eeq
Then the energy loss formula \eqref{DeltaPeq} simplifies to
\beq\label{Elossper}
-\Delta\Big\langle\frac{dE}{dt}\Big\rangle= \sum_{nlm} 4\omega_n^2  Re\left\{   Z_{n lm}^{up *}[J] Z^{down}_{n lm}[J]\, \Scatt_{l}(\omega_n)\right\}\ ,
\eeq
and the ratios \eqref{rattot} and \eqref{ratbh} correspondingly simplify.  In particular, the latter becomes
\beq\label{Cscal}
\ccqo = \sum_{nlm} \omega_n^2 \cdot 2 Re\left\{   Z_{n lm}^{up *}[J] Z^{down}_{n lm}[J]\, \Scatt_{l}(\omega_n)\right\}\Big/\sum_{nlm} \omega_n^2|T_{\omega_n l}|^2 \big |Z^{down}_{nlm}[J]\big|^2\ ,
\eeq
where we have used \eqref{EBHhor}, \eqref{Eemper} to determine the denominator.

\subsection{Orbiting bodies and the circular case}
\label{circsec}

We next apply the preceding formalism to the example of a source corresponding to a body executing an orbit in the background spacetime.  In the limit where the orbiting object can be treated as a pointlike scalar  charge $q$, the corresponding source is
\beq
J=q\int d\tau \frac{\delta^4(x-x(\tau))}{\sqrt{|g|}}\ ,
\eeq
where $x(\tau)$ is the orbital trajectory, and $\tau$ is the body's proper time.  Parameterizing the orbit in terms of Schwarzschild time, this becomes
\beq
J(x)= \frac{q}{r^2 \sin\theta} \frac{d\tau}{dt} \delta(r-r(t)) \delta(\theta-\theta(t)) \delta(\phi-\phi(t)) \ .
\eeq

A particular illustrative example is that of a circular orbit of radius $r_0$ and frequency $\omega_0$, which we take to be in the plane $\theta=\pi/2$, giving
\beq
J(x)= \frac{q}{r_0^2} \frac{d\tau}{dt}\Big\vert_{r_0} \delta(r-r_0) \delta(\theta-\pi/2) \delta(\phi-\omega_0 t)\ .
\eeq
In terms of the relevant wavefunctions (up, down), this then yields, from the definitions \eqref{Zdefs},
\beq
Z_{\omega l m}[J] = \frac{q}{u^t(r_0)} \frac{u^*_{\omega l}(r_0)}{2i\omega r_0} Y^*_{lm}(\pi/2,0) \delta(\omega-m \omega_0) =\sum_n \frac{q}{u^t(r_0)}  \frac{u^*_{n\omega_0 l}(r_0)}{2in\omega_0 r_0} Y^*_{lm}(\pi/2,0) \delta_{mn} \delta(\omega-n\omega_0)
\eeq
where $u^t= dt/d\tau$, and $\omega>0$.

These yield a change in energy loss from \eqref{Elossper}, 
\beq\label{eqn:deltadEdtscalar}
-\Delta\Big\langle\frac{dE}{dt}\Big\rangle = \frac{q^2}{ r_0^2 [u^t(r_0)]^2} \sum_{nl}|Y_{ln}(\pi/2,0)|^2 Re \left[ u_{n\omega_0,l}^{up\, 2}(r_0) \Scatt_l(n\omega_0)\right]
\eeq
and a fractional change compared to the energy absorbed by a classical BH
\beq\label{Ccqo}
\ccqo(\omega_0)=2\frac{\sum_{nl}|Y_{ln}(\pi/2,0)|^2 Re \left[ u_{n\omega_0,l}^{up\, 2}(r_0) \Scatt_l(n\omega_0)\right]}{\sum_{nl}|Y_{ln}(\pi/2,0)|^2 |T_{n\omega_0,l}|^2  |u_{n\omega_0,l}^{up}(r_0)|^2}\ .
\eeq

This can then be computed, given a model for the departures $\Scatt_l(\omega)$ from classical BH scattering, in terms of known BH wavefunctions and transmission coefficients.

\section{Gravitational radiation}\label{sec:grav}

The treatment of gravitational radiation is directly analogous in structure to that of scalar radiation, with the additional complications of the tensor polarizations.  The starting point is the expansion of Einstein's equations
\beq
G_{\mu\nu} = 8\pi G T_{\mu\nu} \, ,
\eeq
in a perturbation $h_{\mu \nu}$ about the BH background.  Solution of the linearized equations is possible after choosing a gauge; for example, we can define the trace-reversed metric perturbation $H_{\mu\nu}$ and impose the Lorenz gauge,
\begin{equation}
H_{\mu\nu} = h_{\mu\nu}-\frac{1}{2}\left(g^{\lambda\sigma}h_{\lambda\sigma}\right)g_{\mu\nu} \, , \quad 	\nabla^{\mu}H_{\mu}{}_{\nu} =0 \, ,
\end{equation} 
resulting in the linearized equation
\begin{equation}\label{eqn:heom}
\left(\square^{\rm (L)}{} H\right)^{\mu\nu}  = \square H^{\mu\nu} + 2 R_{\lambda}{}^{\mu}{}_{\sigma}{}^{\nu}H^{\lambda\sigma} = -16\pi G T^{\mu\nu}(x) \, .
\end{equation} 
This equation, with Lichnerowicz Laplacian $\square^{\rm (L)}$, is the tensor analog to the Klein-Gordon equation \eqref{eqn:KG}.

\subsection{Modes and fluxes}

Eqs.~\eqref{eqn:heom} are no longer in general separable, requiring greater care; the polarization structure, particularly in a rotating background, is complicated.  However, we can define modes using the asymptotics as $r\rightarrow\infty$, where \eqref{eqn:heom} reduces to the flat space homogeneous equation, and we can consider transverse traceless vacuum solutions.  For example, we can define up modes of definite frequency, which are pure outgoing at infinity,
\beq
H^{\rm up}_{\omega A \, \mu \nu}(x) \sim   \frac{e^{-i \omega (t-r)}}{r}  Y_{A \, \mu \nu}\ ;
\eeq
here $A$ is a mode label characterizing the asymptotic behavior of the solution, and $Y_{A\mu \nu }$ are tensor harmonics.\footnote{As this would lead us too far, we do not go into subtleties such as how to deal with non-propagating modes, {\it etc.}}  For example, one may label the modes by the total asymptotic angular quantum numbers, $A=(jm\pm)$, and the $Y_A$'s will be the tensor spherical harmonics.  
We will not need to describe these in detail, however, it is important to note these tensors can be chosen to be orthonormal (with respect to the relevant two-sphere measure), transverse, and trace-free. Moreover,  we have defined these harmonics such that $Y^{(s)}_{\alpha\beta} \sim r^2$, with $\alpha$, $\beta$ labeling angular components. Therefore, the overall scaling is consistent with the expected radiative asymptotic fall-off, which in Cartesian coordinates is $h \sim 1/r$. As a result, in spherical coordinates one has $h_{ab} \sim 1/r$, $h_{\alpha b} \sim 1$ and $h_{\alpha\beta} \sim r$, if $a,b$ run over $t,r$.  One likewise defines down modes to be pure ingoing at infinity, 
\beq\label{eqn:Hdown}
H^{\rm down}_{\omega A \, \mu \nu}(x) \sim  \frac{e^{-i \omega (t+r)}}{r}  Y_{A \, \mu \nu} \, .
\eeq

We fix the normalization of these modes in terms of their asymptotic energy fluxes.  Of course there is no local stress tensor for dynamical gravity.  However, one can still construct a useful conserved energy current $\tau_\mu$ for a massless spin-two field on a fixed background with a time-like Killing vector field, {\it e.g.} as a N\"other current, whose asymptotics \emph{can} meaningfully be interpreted in terms of a physical energy-flux \cite{Sorkin:1991bw, Iyer:1994ys}. This can for example be found from a pseudo stress energy tensor as $\tau_\mu=\tau_{\mu\nu}\xi^\nu$, with possible ambiguities not affecting the physical flux at infinity; we leave the precise choice of $\tau_{\mu\nu}$ at finite $r$ implicit (but see for instance \cite{chandrasekhar1991einstein,Ferrari:2011rb}).
Then we choose norms so that the asymptotic flux of the real or imaginary part of a mode is given by the time-averaged value
\begin{equation}\label{eqn:Ttrnorm}
\lim_{r \to \infty}  \int_{S_2} \, \left\langle \tau^r(H_{\omega A} + H^*_{\omega A}) \right\rangle\,   r^2  d\Omega = \lim_{r \to \infty}  \int_{S_2} \,  -\frac{1}{8 \pi G} 	 \left\langle G^{(2)}_{tr} \right\rangle  r^2  d\Omega = \mp \omega^2 \, ,
\end{equation}
with the sign depending on the choice of ``up'' or ``down'', and where we can  express the asymptotic flux in terms of the second order piece of the Einstein tensor.    (The time averaging also helps
provide a well-defined notion of stress-energy for gravitational waves \cite{Misner:1973prb,Weinberg:1972kfs}.)

As with scalars, $\tau_{\mu\nu}$ can be extended to a symmetric bilinear, and \eqref{eqn:Ttrnorm} generalizes to
\begin{equation}\label{Hupnorm}
\lim_{r \to \infty}  \int_{S_2} \,   \tau^{r}\left( H^{\rm up}_{\omega A},H^{\rm up*}_{\omega A'}\right)  r^2  d\Omega = -\frac{1}{2}\omega^2 \delta_{AA'}  \, ,
\end{equation}
and
\begin{equation}
\lim_{r \to \infty}  \int_{S_2} \,   \tau^{ r}\left( H^{\rm down}_{\omega A},H^{\rm down*}_{\omega A'}\right)  r^2  d\Omega = \frac{1}{2}\omega^2 \delta_{AA'}  \, .
\end{equation}

We also need to evaluate analogous expressions with different modes as arguments.  For example, analogous to the scalar case, we have
\begin{equation}\label{eqn:Trtmixed}
\lim_{r \to \infty}  \int_{{S}_2} \,  \tau^{r}\left( H^{\rm down}_{\omega A}, H^{\rm up*}_{\omega A'} \right)   r^2  d\Omega = 
\lim_{r \to \infty}  \int_{{S}_2} \,  \tau^{r}\left( H^{\rm up}_{\omega A}, H^{\rm down*}_{\omega A'} \right)   r^2  d\Omega=0\ .
\end{equation}
One can check this by explicit computation from \eqref{eqn:Ttrnorm}. Alternatively, one can observe that the ``up'' and ``down'' modes are  related to each other by time-reversal together with complex conjugation, taking $t \to -t$ and $\omega \to - \omega$. This flips the sign of the energy-momentum flux. In particular
\begin{equation}
\lim_{r \to \infty}  \int_{{S}_2} \,  \tau^{ r}\left( H^{\rm down}_{\omega A'}, H^{\rm up*}_{\omega A} \right)    r^2  d\Omega = -
\lim_{r \to \infty}  \int_{{S}_2} \,  \tau^{ r}\left( H^{\rm up *}_{\omega A'}, H^{\rm down}_{\omega A} \right)     r^2  d\Omega\ .
\end{equation}
On the other hand, from symmetry and the asymptotic separation of variables
\begin{equation}
\lim_{r \to \infty}  \int_{{S}_2} \,     \tau^{ r}\left( H^{\rm down}_{\omega A'}, H^{\rm up*}_{\omega A} \right)    r^2  d\Omega = 
\lim_{r \to \infty}  \int_{{S}_2} \,     \tau^{ r}\left( H^{\rm down}_{\omega A}, H^{\rm up*}_{\omega A'} \right)   r^2  d\Omega\ .
\end{equation}
Together this implies \eqref{eqn:Trtmixed}. 

As with the scalar case, expressions with unequal frequencies are also needed for our calculations of the energy fluxes.  Here, too, time averaging is needed for the key relations.  For example, with time averaging as in the scalar case, and as is described further in Appendix \ref{app:averaging}, we have for $\omega\neq\omega'$
\begin{equation}
\lim_{r \to \infty}  \int_{S_2} \, \left\langle  \tau^{r}\left( H^{\rm up}_{\omega A},H^{\rm up*}_{\omega' A'}\right) \right\rangle r^2  d\Omega = 0  \, ,
\end{equation}
and analogous expressions for other unequal-frequency modes.

It is also useful to define the analog of the scalar in and out modes.  These can be defined in terms of the up and down modes, analogously to \eqref{indownreln}, as
\beq\label{tensin}
H^{in}_{\omega A \, \mu\nu}(x)= H^{down}_{\omega A \, \mu\nu}(x) + \sum_{A'} R^{bh}_{AA'}(\omega) H^{up}_{\omega A' \, \mu\nu}(x)\ ;
\eeq
that is, these modes have unit-amplitude (in our normalization \eqref{eqn:Ttrnorm}) incoming wave, and an outgoing reflected wave fixed by the BH boundary conditions at the horizon.  Modes $H^{out}_{\mu\nu\omega A}(x)$ are then related to these by time reversal, complex conjugation, and a corresponding relabelling of indices.

\subsection{Modified scattering, waveforms, and energy loss}

With these preliminaries we can now parameterize the scattering modifications to the classical BH case, and their resulting energy fluxes.  Interactions near the BH can alter the scattered signal, modifying \eqref{tensin} to (compare \eqref{scattbc})
\begin{equation}\label{eqn:gravsc}
H^{\rm sc}_{\omega A \, \mu \nu}(x) = H_{\omega A \, \mu \nu}^{in}(x) +  \int_0^{\infty} \frac{d\omega'}{\sqrt{2 \omega'}} \sum_{A'} \Scatt_{A A'} (\omega,\omega')
 H^{\rm up}_{\omega' A' \, \mu \nu}(x)  \, ;
\end{equation}
the incoming unit-amplitude signal scatters into different outgoing waves, not necessarily preserving $A'$.

For a given source $T_{\mu\nu}$, the signal with modified scattering is expected to be determined by a Green function,
\beq\label{BHgreenH}
H^{\mu \nu}_T(x)= \int dV_4' \, G^{\mu \nu,\lambda\sigma}(x,x') T_{\lambda\sigma}(x') \, .
\eeq 
Outside the scattering region, $r>R_a$, the difference between any two Green functions 
\beq\label{BHGdiff}
G_{\mu \nu,\lambda\sigma}^{sc}(x,x')- G_{\mu \nu,\lambda\sigma}^{bh}(x,x') = \Delta G_{\mu \nu,\lambda\sigma}(x,x')\ ,
\eeq
is again a solution of the homogeneous equation in $x$.  
Consider the BH Green function for $r<r'$, where it is a solution of the homogeneous equation, and thus may be expanded in terms of the solutions $H^{in}$ (enforcing the BH boundary conditions),
\beq\label{GBHT}
G_{\mu \nu,\lambda\sigma}^{bh}(x,x') =\sum_A\int_0^\infty d\omega K^{<\omega A}_{\lambda\sigma}(x') H^{in}_{\omega A \, \mu\nu}(x) + cc\ ,
\eeq
with coefficient functions $K^<$ in principle determinable by a matching procedure like the scalar case.  Then, the boundary conditions corresponding to scattering modifications \eqref{eqn:gravsc} of the CQO lead to the modification
\beq\label{BHmod}
\Delta G_{\mu \nu,\lambda\sigma}=\sum_A \int_0^\infty d\omega K^{<\omega A}_{\lambda\sigma}(x') \int_0^{\infty} \frac{d\omega'}{\sqrt{2 \omega'}} \sum_{A'} \Scatt_{AA'}(\omega,\omega')
 H^{\rm up}_{\omega' A' \, \mu \nu}(x) + cc \, ,
\eeq
to the Green function, as in \eqref{PerturbedG}.  As in that case, since $\Delta G$ is a homogeneous solution, we once again expect this expression to extend to $r>r'$.  For $r>r'$, we also expect an expression of the form
\beq\label{Gtensout}
G_{\mu \nu,\lambda\sigma}^{bh}(x,x') =\sum_A\int_0^\infty d\omega K^{>\omega A}_{\lambda\sigma}(x') H^{up}_{\omega A \, \mu\nu}(x) + cc\ ,
\eeq
with coefficient functions $K^>$ again determined by a matching procedure.

The source $T_{\mu\nu}$ produces a perturbed solution
\beq
H^T(x) = H^{bh\, T}(x) + \Delta H^T(x)\ ,
\eeq
analogous to the scalar \eqref{newsoln}.  From \eqref{BHgreenH}, \eqref{BHGdiff}, and \eqref{BHmod}, the perturbation tensor signal is
\beq\label{DeltaHT}
\Delta H^T_{\mu\nu} = \sum_{AA'} \int_0^{\infty} \frac{d\omega d\omega'}{\sqrt{2\omega'}} Z^{down}_{\omega A}[T] \Scatt_{AA'}(\omega,\omega')
 H^{\rm up}_{\omega' A' \, \mu \nu}(x) + cc \, ,  
 \eeq
where we define
\beq
Z^{down}_{\omega A}[T] = \int dV_4 K^{<\omega A}_{\lambda\sigma}(x)\, T^{\lambda\sigma}(x)\ .
\eeq
We likewise define, using the scalar analogy,
\beq
Z^{out}_{\omega A}[T] =  \int dV_4 K^{>\omega A}_{\lambda\sigma}(x)\, T^{\lambda\sigma}(x)\ .
\eeq

Eq.~\eqref{DeltaHT} gives us a prescription to calculate the change in the gravitational wave signal due to a given source such as an orbiting body, in terms of the parameters $\Scatt_{AA'}(\omega,\omega')$ that we have introduced to describe modifications to scattering.  In principle the signal deviation could be directly measurable, but in practice this may be difficult if $\Scatt$ and the resulting change in signal is small.  However, these changes in the waveform will also generically result in a change in the rate at which energy is emitted, and over many orbits this can accumulate to enhance the significance of the corrections in their contribution to the phase.  We therefore turn to the question of the change in emitted energy.

The argument for the modification to the energy loss also follows that of the scalar case.
The energy flux of the perturbed signal is described by
\beq\label{tenspert}
\tau_{\mu}(H^T, H^T)= \tau_{\mu}(H^{bh\, T}, H^{bh\, T}) + \tau_{\mu}(\Delta H^T, \Delta H^T) + 2 \tau_{\mu}(H^{bh\, T}, \Delta H^T)\ .
\eeq
Again let the source $T_{\mu\nu}$ be bounded between radii $r_1<r_2$; this is valid at leading order in perturbation theory in $G$, but of course is violated at higher order.  Since $\Delta H^T$ is a solution of the homogeneous equations in this region, the second term in \eqref{tenspert} is conserved in steady state (or when time averaged) and doesn't contribute to the total power emitted by the source,
\beq\label{Totpower}
-\frac{dE}{dt} = -\left(\int_{r_2} \rho^2 d\Omega - \int_{r_1} \rho^2 d\Omega\right) \tau^r (H^T,H^T) \ 
\eeq
where we have introduced $\rho^2=r^2+a^2\cos^2\theta$ to generalize to the rotating case with $a=S/M$.
Again the change in power emitted due to the CQO boundary conditions is linear in $\Scatt$, and given by 
\beq\label{Empowermet}
-\Delta \frac{dE}{dt} = -2\left(\int_{r_2} \rho^2 d\Omega - \int_{r_1} \rho^2 d\Omega\right) \tau^r(H^{bh\,T},\Delta H^T)\ .
\eeq

We evaluate this power as in the scalar case.  For $r<r_1$, we see from the Green function form \eqref{GBHT} and the solutions \eqref{tensin} that 
 the BH background solution $H^{bh\,T}$ has a direct down piece, superposing $H^{down}_{\omega A}$'s, and a reflected up piece, superposing $H^{up}_{\omega A}$'s.
 For $r>r_2$,  we see from \eqref{Gtensout} and \eqref{BHgreenH} that the up signal may be decomposed into the continuation of the up signal from $r<r_1$, plus a direct up signal (compare \eqref{phidirJ}),
 \beq\label{Hbhdir}
H^{bh,>direct\, T}_{\mu\nu} = \sum_A \int_0^\infty d\omega Z^{up}_{\omega A}[T] H^{up}_{\omega A \, \mu\nu} + cc\ ,
\eeq
where for present purposes we define
\beq
Z^{up}_{\omega A}[T]= Z^{out}_{\omega A}[T] -\sum_{A'} Z^{down}_{\omega A'}[T] R^{bh}_{A'A}\ .
\eeq
This is the tensor version of \eqref{eqn:Zrel}.

Next note that if $H_1$ and $H_2$ are homogeneous solutions, then
\beq\label{constwo}
\nabla^\mu \tau_\mu(H_1, H_2)=0
\eeq
follows from vanishing of $\nabla^\mu \tau_\mu(H_1+H_2, H_1+H_2)$.  Thus, after averaging $\langle \tau_{r}(H^{bh,reflected,T},\Delta H^T)\rangle$ is conserved, and doesn't contribute to the difference between $r_1$ and $r_2$ in \eqref{Empowermet}.  Also, as
 in the scalar case, the contribution from $\tau_{r}(H^{bh,direct\,T},\Delta H^T)$ at $r_1$ vanishes under time averaging, since it involves $\tau_{r}(H^{down},H^{up})$, $\tau_{r}(H^{down},H^{up*})$ or their complex conjugates.  Time averaging eliminates the first, and projects the second on equal frequencies, where it is time independent.  Then conservation \eqref{constwo} applied to the integral $\int \rho^2 d\Omega \langle\tau^r(H^{down},H^{up*})\rangle$ relates it to its value at infinity, where it vanishes by \eqref{eqn:Trtmixed}.
As a result the change in average radiated power is simply
\beq
-\Delta\Big\langle\frac{dE}{dt}\Big\rangle = -2 \int_{r_2} \rho^2 d\Omega \langle \tau^r(H^{bh, >direct\, T}, \Delta H^T)\rangle\ .
\eeq
Combining this with the expressions \eqref{Hbhdir} and \eqref{DeltaHT} then gives the result (compare \eqref{DeltaP})
\beq 
-\Delta \left\langle \frac{dE}{dt}\right\rangle =  2\sum_{AA'}\int_0^\infty d\omega  d\omega' \frac{d\omega''}{\sqrt{2\omega''}} \omega^2 Re\left\{\Delta_T(\omega-\omega'') Z^{up*}_{\omega A}[T] Z^{down}_{\omega'A'}\Scatt_{A'A}(\omega',\omega'')\right\} \ ,
\eeq
where we have used the normalization \eqref{Hupnorm}, and time averaging enters through $\Delta_T(\omega-\omega'')$, which sets the frequencies equal, as in the scalar case (see Section \ref{sec:energyloss} and Appendix \ref{app:averaging} for more discussion).

As in the scalar case, we can decompose $\Scatt$ into equal and unequal frequency cases,
\beq
\Scatt_{A'A}(\omega',\omega)= \sqrt{2\omega} \Scatt_{A'A}(\omega)\delta(\omega-\omega') + \Scatt_{A'A}^{\neq}(\omega',\omega) 
\eeq
If we focus on the equal frequency case, the emitted power becomes
\beq\label{DeltaPeqT}
-\Delta\Big\langle\frac{dE}{dt}\Big\rangle = 2\sum_{AA'} \int_0^\infty d\omega d\omega'  \omega^2  Re\left\{ \Delta_T(\omega-\omega')  Z_{\omega A}^{up *}[T] Z^{down}_{\omega' A'}[T]\, \Scatt_{A'A}(\omega')\right\} \ .
\eeq
We can describe its relative effect by comparing this to the power absorbed by the would-be classical BH, again with the definition \eqref{ratbh}.  In the special case where $T_{\mu\nu}$ has discrete frequencies, as with a periodic orbit,
\beq
T_{\mu\nu}= \sum_{n=0}^\infty T_{n\mu\nu}({\vec x}) e^{-i\omega_n t} + cc \, ,
\eeq
the energy loss formula reduces to
\beq
-\Delta\Big\langle\frac{dE}{dt}\Big\rangle= \sum_{nAA'} \omega_n^2 \cdot 2 Re\left\{   Z_{n A}^{up *}[T] Z^{down}_{n A'}[T]\, \Scatt_{A'A}(\omega_n)\right\}\ , 
\eeq
where
\beq
Z_{\omega A}=\sum_n Z_{nA}[T] \delta(\omega-\omega_n)\ .
\eeq
We then find the ratio
\beq\label{Ctens}
\ccqo = \sum_{nAA'} \omega_n^2 \cdot 2 Re\left\{   Z_{n A}^{up *}[T] Z^{down}_{nA'}[T]\, \Scatt_{A'A}(\omega_n)\right\}\Big/\sum_{nAA'} \omega_n^2|T_{\omega_n AA'}|^2 Z^{down}_{nA}[T]  Z^{down*}_{nA'}[T] \ ,
\eeq
where we have written the energy flux expression into the BH in terms of an effective transmission coefficient defined in terms of the time average and angular integral of $\tau_{r}(H^{in*}_A,H^{in}_{A'})$ at the horizon.  

The direct comparison of the expressions for $\ccqo$ in the scalar and tensor cases, exemplified by comparing \eqref{Cscal} and \eqref{Ctens}, indicate that the scalar versions, which are simpler to compute and handle, serve as a useful proxy for the more complicated tensor case.  Specifically, we find that the ratio $\ccqo$ involves closely similar ratios of wavefunction factors integrated against sources, with source strengths that cancel in the ratio, and similar appearance of the parameters $\Scatt$ describing the CQO perturbation to classical BH scattering.  The primary difference in the tensor case is the more complicated polarization structure.  We do expect this to introduce extra factors into the ratios, but expect these to be order one factors.  For purposes of understanding the sensitivity of gravitational wave signatures to the scattering perturbations at an order-of-magnitude level, this therefore motivates working with the much simpler scalar models as a simple example.
Of course, a fully accurate calculation must take into account the full tensor structure, {\it e.g.}, as described in this section.

\subsection{Teukolsky variables}

In order to translate the analysis from metric perturbations to Teukolsky variables, we need to compute the associated linearized (Weyl) curvature perturbations $\delta C^{\mu}{}_{\nu \alpha \beta}[H;g]$ and project onto a principal null frame \cite{Teukolsky:1973ha}
\begin{equation}
\psi_4[H] = \psi_4[h;g, \left\lbrace  l^{\mu},n^{\mu},m^{\mu},\bar{m}^{\mu} \right\rbrace] =  n_{\mu} \bar{m}^{\nu} n^{\alpha} \bar{m}^{\beta}\delta C^{\mu}{}_{\nu \alpha \beta}[H;g]\, ,
\end{equation}
indicating with the first equality explicitly all the background input: the background metric $g$ and principal null frame $\left\lbrace l^{\mu},n^{\mu},m^{\mu},\bar{m}^{\mu} \right\rbrace$. We could have also introduced $\psi_0[H]$ by simply replacing $n^{\mu} \rightarrow l^{\mu}$, $\bar m^{\mu} \rightarrow m^{\mu}$. \\

By linearity, if we define
\begin{equation}
\psi^{sc}_{\omega A}	= \psi_4[H^{sc}_{\omega A}] \, , \quad  \psi^{in}_{\omega A} =	\psi_4[H^{in}_{\omega A}] \, , \quad  \psi^{up}_{\omega A}=	\psi_4[H^{up}_{\omega A}] \, 
\end{equation}
(or similarly for $\psi_0$)
then \eqref{eqn:gravsc} becomes
\begin{equation}\label{eqn:gravsct}
\psi^{sc}_{\omega A} = \psi^{in}_{\omega A} +  \int_0^{\infty} \frac{d\omega'}{\sqrt{2 \omega'}} \sum_{A'} \Scatt_{A A'}(\omega, \omega')
\psi^{up}_{\omega' A'}  \, .
\end{equation}
On the other hand, $\psi^{in}_{\omega A}$ and  $\psi^{up}_{\omega A}$ would generically not be the separated, single mode solutions to the Teukolsky equation \cite{Teukolsky:1973ha}. 
If $\Psi^{in}_{\omega b}$, $\Psi^{up}_{\omega b}$ are a basis of such single mode solutions with the appropriate boundary conditions, we can write
\begin{equation}
\Psi^{in}_{\omega b} = \sum_A M^{-1}_{b A}(\omega a)\psi^{in}_{\omega A}  \, , \quad 	\psi^{up}_{\omega A} = \sum_b N_{A b}(\omega a) \Psi^{up}_{\omega b} \, 
\end{equation}
with $a=S/M$.
$M^{-1}_{b A}(\omega a)$ and $N_{Ab}(\omega a)$  encode the change of basis, which are frequency and spin dependent, as are the angular Teukolsky mode functions. A subtlety with the former is that, even if $H^{in}_{\omega A}$ is a proper basis for the ingoing modes, $\psi^{in}_{\omega A}$ may be overcomplete. For instance, one could have metric perturbations related by ``completion'' pieces of \cite{Merlin:2016boc, Toomani:2021jlo}. Those however, would correspond to non-propagating modes alluded to earlier. Then, defining $\Scatt_{ bb'}(\omega,\omega') $ as
\begin{equation}\label{eqn:gravscTeukosly}
\Psi^{sc}_{\omega b} = \Psi^{in}_{\omega b} +  \int_0^{\infty} \frac{d\omega'}{\sqrt{2 \omega'}} \sum_{b'} \Scatt_{b b'}(\omega, \omega')
\Psi^{up}_{\omega' b'}  \, .
\end{equation}
one has the change of basis relation
\begin{equation}
\Scatt_{b b'}(\omega, \omega') = \sum_A\sum_{A'} M^{-1}_{bA}(\omega a) \Scatt_{AA'}(\omega,\omega') N_{A' b'}(\omega' a) \, ,
\end{equation}
connecting the parameterization in terms of metric perturbations to that in terms of Teukolsky variables.

\section{Relation to models for scattering}\label{sec:comparison}

We have given in \eqref{phiscatt} and \eqref{hscatt} a very general parameterization of modifications to scattering from a would-be black hole due to departures from the standard classical black hole geometry.  An important question is what such corrections a given underlying detailed model of the physics produces.  For present purposes of illustration we consider only some particularly simple models, which involve new scattering contributions in the vicinity of the would-be horizon, which can be parameterized in terms of effective reflection coefficients.  Other models that exist in the literature include modification of the Regge-Wheeler potential \cite{Volkel:2017kfj,Volkel:2019ahb,Volkel:2019gpq}.  And, if the scattering arises from new interactions associated with restoration of unitarity, such as in nonviolent unitarization\cite{Giddings:2011ks,Giddings:2012gc,Giddings:2013kcj,Giddings:2014nla,Giddings:2017mym,Giddings:2019vvj,Giddings:2022ipt}, then the scattering might for example be described as arising from interaction terms in an effective hamiltonian.  From the point of view of the asymptotic observer, the scattering amplitudes are the more directly physically accessible quantities.  Moreover, related phenomena such as modifications to Love numbers \cite{Cardoso:2017cfl} and quasinormal modes \cite{Cardoso:2019mqo}, can be directly encoded in such amplitudes.

\subsection{Near-horizon boundary conditions}
\label{NHBCsec}

In recent literature, a common model for modifications to classical BH behavior assumes the existence of a modified boundary condition for scattering very close to the would-be horizon\cite{Abedi:2016hgu,Datta:2019epe,Maggio:2020jml,Maggio:2021uge,Sago:2021iku,Sago:2022bbj}.  This can be heuristically motivated by comparing to physically realized systems like neutron stars\cite{Allen:1997xj,Kokkotas:1999bd,Kokkotas:2000up}.  However, one challenge for this approach is to give a physical realization of a matter or other classical configuration that only departs from the BH vacuum geometry very near the horizon.  For example, a fluid with positive pressure and energy density gives solutions satisfying the Buchdahl bound\cite{Buchdahl:1956zz}, giving radius $R_c>9M/4$; enforcing causal propagation of sound, at a speed less than that of light, gives the more restrictive \cite{GKT} $R_c>2.82 M$.  This type of model does however serve as a useful illustration of our more general approach to parameterizing deviations from classical BH behavior.


Specifically consider imposing a boundary condition at a radius $r=(1+\epsilon) R$ in the Schwarzschild geometry, with $\epsilon\ll 1$.  This value of $r$ will correspond to a value $r_{*} \ll -R$.  Here the effective potential \eqref{RWpot} asymptotes to zero, and BH boundary conditions for the in modes are given in \eqref{eqn:Rinoutdefsscal}.  These are modified by assuming nonzero reflection at $r_*=r_{*\epsilon}$, which in general may be taken to depend on frequency and angular momentum.  For $r_{*\epsilon}\ll -R$, where the solutions are well-approximated as plane waves in $r_*$, a good approximate form of the reflecting boundary condition arises by assuming that near $r_*=r_{*\epsilon}$,
\beq
u_{\omega l}(r) \propto e^{-i\omega r_*} + \hat \calr_{\omega l} e^{i\omega(r_*-2r_{*\epsilon})}\ ,
\eeq
with reflection coefficient $\hat \calr_{\omega l}$.  This is also well approximated near $r_*=r_{*\epsilon}$ by taking the solutions to be of the form
\beq\label{NHBCdef}
u_{\omega l}(r)\propto u_{\omega l}^{in}(r) + \hat \calr_{\omega l} e^{-2 i \omega r_{*\epsilon}} u_{\omega l}^{in*}(r)\quad ;
\eeq
to simplify, define 
\beq 
\tilde \calr_{\omega l}=\hat \calr_{\omega l} e^{-2 i \omega r_{*\epsilon}}\ .
\eeq

The boundary condition may be written in the form described in Sec.~\ref{sec:energyloss} by first using the relations \eqref{indownreln} and \eqref{ccreln} to write
\beq
u_{\omega l}^{out}(r) = u_{\omega l}^{in*}(r) = \left(1-|R_{\omega l}|^2\right) u_{\omega l}^{up}(r) + R_{\omega l}^* u_{\omega l}^{in}(r)= |T_{\omega l}|^2 u_{\omega l}^{up}(r) + R_{\omega l}^* u_{\omega l}^{in}(r)
\eeq
where the last equality uses $|R_{\omega l}|^2+ |T_{\omega l}|^2=1$, which follows from constancy of the Wronskian $W_{r_*}$.
Then, \eqref{NHBCdef} becomes
\beq
u_{\omega l}(r) \propto  u_{\omega l}^{in} \left(1+ {\tilde \calr_{\omega l}} R_{\omega l}^*\right) + {\tilde \calr_{\omega l}} |T_{\omega l}|^2 u_{\omega l}^{up} \ , 
\eeq
or, normalizing to unit amplitude in at $r_*=\infty$, 
\beq
u_{\omega l} = u_{\omega l}^{in} + \frac{ {\tilde \calr_{\omega l}}  |T_{\omega l}|^2}{1+{\tilde \calr_{\omega l}} R_{\omega l}^*} u_{\omega l}^{up}\ .
\eeq
Comparing to our definitions \eqref{scattbc} and \eqref{Tdecomp} gives 
\beq\label{scattnh}
\Scatt_l(\omega)=  \frac{ {\tilde \calr_{\omega l}}  |T_{\omega l}|^2}{1+{\tilde \calr_{\omega l}} R_{\omega l}^*}
\eeq
for models with the near-horizon reflecting boundary conditions \eqref{NHBCdef}.  This may then be used directly in the formula \eqref{DeltaPeq} for the change in the energy loss.

The scattering amplitude \eqref{scattnh} exhibits various notable features.  To interpret the denominator in \eqref{scattnh}, consider expanding it in $\tilde \cR$,
\beq\label{NHBCexp}
\Scatt_l(\omega) = {\tilde \calr_{\omega l}}  |T_{\omega l}|^2 \left[1+ \sum_{k=1}^\infty \left(- \tilde\calr_{\omega l} R_{\omega l}^*\right)^k\right]\ .
\eeq
This has an intuitive interpretation as a sum of contributions from multiple reflections in the near-horizon region, between the reflecting surface and the potential,
followed by transmission.  This directly connects to the phenomenon of ``echoes'' in gravitational wave signals \cite{Cardoso:2016rao,Cardoso:2016oxy}\cite{Abedi:2016hgu}, which is therefore incorporated in this analysis.  Of course, for small $\tilde \calr$ and away from zeroes of the denominator of \eqref{scattnh}, this reduces to the linear result 
\beq\label{linapprox}
\Scatt_l(\omega) \approx {\tilde \calr_{\omega l}}  |T_{\omega l}|^2\ .
\eeq

A related observation is that, from \eqref{linapprox}, we find that to leading order the change in the energy loss 
\eqref{DeltaPeq}, \eqref{DeltaPeqT}  is linear in  $\tilde \cR$.  This contrasts with the assumptions of \cite{Datta:2019epe} (see {\it e.g.} their eq. (12)) that the change in energy loss is quadratic in such a reflection coefficient.  This doesn't have major consequences for their analysis, which is effectively giving bounds on $\ccqo$ (see the next section), but rather affects the interpretation of the parameter they call $\cal R$.  

In addition to the echo phenomenon, an $\epsilon\ll1$ near-horizon boundary condition also gives rise to a related universal family of ``trapped'' quasinormal modes.  These correspond to zeroes of the denominator of \eqref{scattnh}, so occur at 
\beq
{\tilde \calr_{\omega l}} R_{\omega l}^*=-1\ .
\eeq
For $r_{*\epsilon}\rightarrow-\infty$, these are low-frequency modes which can in principle be excited in the binary inspiral. Following related work for neutron star binaries \cite{Lai:1993di,Lai:1997wh,Tsang:2011ad}, such excitations have been the subject of separate studies \cite{Cardoso:2019nis,Cardoso:2022fbq}.  These studies indicate that, while potentially interesting, they often present themselves as unobservable ``glitches'' in the gravitational wave signal.

Ref.~\cite{Maggio:2021uge} extends the analysis of \cite{Datta:2019epe} to derive corresponding bounds on the reflection parameters $\tilde \calr_{\omega l}$.  We will discuss such bounds in the next section, but note a few aspects of that discussion here.  

A simplest model is frequency-independent  $\tilde \calr_{\omega l}$.  This clearly exhibits the resonant phenomena we have just described.  In addition, note that at high frequencies, these crude models significantly change the entire scattering, as parameterized by $\Scatt$.  This arises since the  transmission coefficients $T_{\omega l}$ become unity at large $\omega$.  However, such frequencies are not effectively probed by energy loss during quasicircular  inspiral, as is seen {\it e.g.} from \eqref{Ccqo} in which the maximum frequency $l\omega_0$ enters (see further discussion in the next section).  At such low frequencies, the transmission coefficients lead to significant suppression; in this context, one should not overemphasize that the classical general relativistic black hole is a perfect absorber as there is still a significant dynamical barrier separating one from the horizon where this is strictly true. 

An alternate model for frequency dependence assumes Boltzmann behavior\cite{Oshita:2019sat,Wang:2019rcf}, with parameters determined by Hawking temperature $T_H$ and angular velocity $\Omega_H$ of the classical BH solution with the same parameters as the underlying quantum object,
\begin{equation}\label{eqn:boltzmannreflectivity}
\tilde\calr_{\omega l m}^{\rm Boltzmann} = \calr e^{- \frac{|\omega-m\Omega_H|}{ 2T_H}} \, .
\end{equation}
Such a model suppresses the high-frequency deviation in $\Scatt$, found via \eqref{scattnh}.  
Bounds on such models are also studied in \cite{Maggio:2021uge}, with similar results to the frequency-independent models for slowly rotating black holes. At significant spin, the high-frequency suppression kicks in, even for low orbital frequencies, due to the \emph{relative} high frequency with respect to the black hole rotational frequency. This model thus presents a mechanism \emph{preventing} higher spin from improving the observational prospects.

\subsection{Scattering from a black hole ``quantum halo"}
\label{scatta}

It is a difficult challenge to give a detailed model of the underlying physics in which modifications to the classical black hole geometry only appear a microscopic distance above the horizon.  One can think of this as partly due to a type of naturalness problem; since a natural scale in the problem is the horizon radius $R$, restricting new physics to a region of size $\Delta r\ll R$ requires some  new scale and/or fine tuning in the physics.  Isolating new physics to such a narrow region also has the consequence that such physics is seen to be extremely ``hard" to infalling observers, corresponding to a large violation of the equivalence principle; in the quantum context, this is exemplified in the firewall proposal\cite{AMPS}.

There are strong indications that in order to unitarize black hole quantum evolution, resolving the ``information paradox," some new physics is needed outside the horizon.  However, it appears that this new physics can represent both a less extreme violation of the equivalence principle and have a less violent effect on infalling observers if it occurs at much larger distances than a Planck distance from the horizon; this permits it to be ``softer."  In particular, if new interactions are present on scales separated by $\sim R^p$ from the horizon, with $0<p\leq1$, the violation and the violence become less extreme for larger black holes.  The proposal of ``nonviolent unitarization\cite{Giddings:2011ks,Giddings:2012gc,Giddings:2013kcj,Giddings:2014nla,Giddings:2017mym,Giddings:2019vvj,Giddings:2022ipt}" is that the full quantum description can be parameterized as having  interactions on such scales, in what might be called a ``quantum halo," that are responsible for reinstating unitarity in BH evaporation.  The resulting compact quantum object  has many of the coarse-grained properties of a black hole; scattering from it is  very similar to that of a black hole, but there are deviations associated with the physics restoring unitary quantum evolution.  We will focus on the example  of $p=1$; it is notable that  there are good reasons to think of the Hawking radiation as also produced at such scales\cite{SGBoltz}.\footnote{Recent work of \cite{BoPe} has argued instead for $p=1/2$.}

Microscopic models for such interactions have been considered elsewhere\cite{Giddings:2017mym,Giddings:2019vvj,Giddings:2022ipt}, and will be investigated further in the future, but for present purposes we seek a simple model to explore the possible sensitivity of gravitational wave observations to such interactions.  Such a simple model is provided by assuming that interactions introduce some additional elastic reflection of incident partial waves at a radius $R_a$.  Unit magnitude reflection at $R_a$ can for example be described by imposing Dirichlet boundary conditions on \eqref{efscatt},
\beq
u_{\omega l}^{\rm in}(R_a) +\Scatt_l(\omega)   u_{\omega l}^{\rm up}(R_a)=0\ ,
\eeq
implying
\beq
\Scatt_l(\omega)= -u_{\omega l}^{\rm in}(R_a)/ u_{\omega l}^{\rm up}(R_a)\ .
\eeq
Likewise, partial reflection at $R_a$ may be parameterized by an effective reflection coefficient $ \calr_{\omega l}(R_a)$, analogously defined by
\beq\label{refdef}
\Scatt_l(\omega)=  \calr_{\omega l}(R_a)\, \frac{u_{\omega l}^{\rm in}(R_a)}{ u_{\omega l}^{\rm up}(R_a)}\ .
\eeq
The factors of the wavefunction in the relation between  $\calr_{\omega l}$ and $\Scatt_l(\omega)$ account for tunneling factors to and from the scattering radius $R_a$; this definition is clearly analogous to the relation \eqref{scattnh}, without the explicit multiple reflections which we expect in that case.  This means that $ \calr_{\omega l}(R_a)$ serves as a useful parameterization of the strength of the interaction inducing scattering if it occurs at radius $\sim R_a$; observational bounds on $\Scatt_l(\omega)$ may be converted into bounds on this effective parameter via the relation \eqref{refdef}.

\section{Prospective observational bounds}\label{sec:observational}

For simplicity, we consider bounds arising from circular orbits, as described in \ref{circsec}, and investigated in \cite{Datta:2019epe,Maggio:2021uge,Sago:2021iku}.  From the formulas \eqref{Ctens}, \eqref{Ccqo} for $\ccqo$, we see that the bounds on deviations in energy emission from that of the BH case will provide constraints on the basic scattering parameters $\Scatt_{AA'}(\omega_n)$, or in the simple scalar example,  $\Scatt_l(n\omega_0)$, with $n\leq l$, where the orbital frequency is given by
\beq
\omega_0=\frac{M^{1/2}}{r_0^{3/2}}
\eeq
in terms of the central mass\footnote{The masses and spins of the black holes in the binary will change during the inspiral; for instance by absorption of gravitational radiation. However, we treat them as constant because this change is subleading in our analysis \cite{Hughes:2018qxz}. See also \cite{deCesare:2023rmg}.} $M$ and the orbital radius $r_0$.  The bound on $n$ means that we can constrain scattering for incident waves at frequencies $\omega\leq \omega_0 l$.  
An effective impact parameter for the highest frequency of these waves is $l/\omega = 1/\omega_0=r_0\sqrt{r_0/M}$.  

	In a quasicircular inspiral, the orbital phase increases at a rate set by the instantaneous orbital frequency, whose evolution in turn is set by the energy-flux and the gradient in the orbital energy $E_0$
\beq
\frac{d \phi_0}{dt} = \omega_0 \, , \quad \frac{d \omega_0}{dt} = \left\langle \frac{d E}{dt}\right\rangle \left(\frac{d E_0}{d \omega_0}\right)^{-1} \, .
\eeq
The difference in phase compared to that for a central BH, accumulated from a reference orbital frequency $\omega_{\rm ref}$, is given to leading order by 
\beq
\Delta \phi_0 = -\int^{\omega}_{\omega_{\rm ref}} d\omega_0 \, \omega_0 \left\langle \frac{d E}{dt}\right\rangle_{BH, tot}^{-2}\left(\frac{d E_0}{d \omega_0}\right) \Delta \left\langle \frac{d E}{dt}\right\rangle = \int^{\omega}_{\omega_{\rm ref}} d\omega_0 \frac{d\phi_{BH}}{d\omega_0} \frac{-\Delta \left\langle \frac{d E}{dt}\right\rangle}{\left\langle \frac{d E}{dt}\right\rangle_{BH, tot}}\ ,
\eeq
where $-\left\langle \frac{d E}{dt}\right\rangle_{BH, tot}$ is the total power radiated by the object orbiting a classical BH, and where we have defined the BH phase evolution
\beq
\frac{d\phi_{BH}}{d\omega_0} = \omega_0\Big/\left(\frac{d \omega_0}{dt}\right)_{BH}\ .
\eeq
Instead of normalizing the perturbed energy loss to the total emitted energy, \cite{Datta:2019epe} normalizes it to the energy absorbed into the BH horizon; we have followed this in defining $\ccqo$ in \eqref{ratbh}.  Introducing the ratio
\beq
\rho_{BH,abs}=\left\langle \frac{d E}{dt}\right\rangle_{BH, hor}/\left\langle \frac{d E}{dt}\right\rangle_{BH, tot} \,
\eeq
the dephasing associated to a compact quantum object parameterized by $\ccqo$ is then given by
\beq\label{eqn:phaseshift}
\Delta \phi_0 = -\int^{\omega}_{\omega_{\rm ref}}d\omega_0 \frac{d\phi_{BH}}{d\omega_0}  \, \, \rho_{BH,abs}\, \ccqo \, .
\eeq

\begin{figure}[t!]
			\begin{center}
				\includegraphics[width=0.48\textwidth]{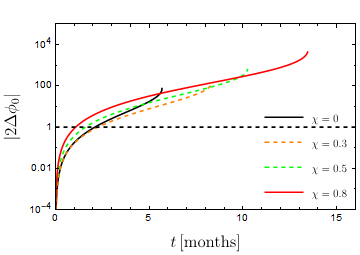} 
				\includegraphics[width=0.48\textwidth]{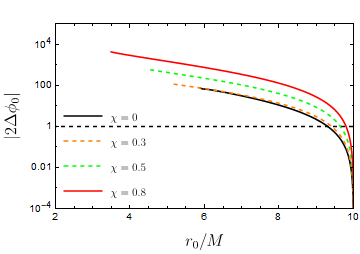} 
				\caption{Accumulating phase difference for an inspiral of a $30M_{\odot}$ black hole into a $M=10^6 M_{\odot}$ supermassive black hole starting at $10M$ for various values of the supermassive black hole spin and constant $\ccqo$ (left) and the same but as a function of orbital radius of the black hole inspiral (right); see also Fig.~2 of \cite{Datta:2019epe}.}
				\label{Fig:HughesPhasesShifts}
			\end{center}
\end{figure}

The analysis of \cite{Datta:2019epe} effectively took the quantity $\ccqo$ to be constant in frequency; as we have described, the relation to an underlying reflection amplitude differs from what \cite{Datta:2019epe} assumed, so for purposes of summarizing their bounds we treat $\ccqo$ as the physical parameter that we are bounding by measurements of the gravitational wave phase.
For a frequency range corresponding to an inspiral into a Schwarzschild BH from $r_0=10M$ to ISCO at $r_0 = 6M$, such a constant $\ccqo$ would give $\Delta \phi_0 \sim  - 10^{-3} \frac{M}{\mu} \ccqo$, with $\mu$ the mass of the lighter inspiraling object, as can be found from numerical evaluation of \eqref{eqn:phaseshift}.  This evaluation is accomplished by using a continued fraction method from the Black Hole Perturbation Toolkit \cite{BHPToolkit} for the energy-fluxes.\footnote{The relevant energy-flux data is also made directly available in the Black Hole Perturbation Toolkit based on \cite{Taracchini:2014zpa} and can be verified against \cite{Finn:2000sy}, who use a Sasaki-Nakamura formulation of Teukolsky equations \cite{Sasaki:1981sx}. See also \cite{Nasipak:2023kuf}.}  The mass ratio dependence arises from  the ratio between the orbital energy, which is proportional to  $\mu$, and the energy-flux $\propto \mu^2$, analogously to how the $q^2$ dependence arises in \eqref{eqn:deltadEdtscalar}. See Fig.~\ref{Fig:HughesPhasesShifts} for more fine-grained results on $\Delta \phi_0$ for constant $\ccqo$; this reproduces (parts of) Fig.~2 of \cite{Datta:2019epe}, but also includes the Schwarzschild case.
		
If one assumes a mass ratio of a supermassive black hole to a stellar mass black hole of about $10^5$ and a detection threshold of about $2\Delta \phi_0 \sim 1$ \cite{Lindblom:2008cm}, the example discussed above would yield an achievable sensitivity $\ccqo \lesssim 10^{-2}$. As long as the full inspiral from $10M$ to the innermost stable orbit is observed, these results are readily scaled for different mass-ratios.

However, rotation can significantly improve such a bound, as was found in \cite{Datta:2019epe}.  There they assumed a dimensionless spin $\chi=S/M^2$ of $\chi \approx 0.8$, as well as using a slightly more sophisticated measure, and estimated one could achieve $\ccqo \lesssim 10^{-4}$. One can see this improvement in sensitivity directly from numerical integration of \eqref{eqn:phaseshift}, which for $\chi=0.8$ yields the  phase shift $2\Delta \phi_0 \approx  10^{-1} \frac{M}{\mu} \ccqo$.  Comparison to the previous result for Schwarzschild directly illustrates the improvement in sensitivity to $\ccqo$ by a factor $\sim 10^2$.  This enhancement can be understood as being due to a combination of the larger relative horizon absorption at a given (corotating) orbital radius, as well as to the smaller innermost stable orbit; see Fig.~\ref{Fig:HughesPhasesShifts} where the phase shifts are  higher early on due to the former effect and last longer because of the latter.

Similar results are found in simple models based on a direct implementation of near-horizon reflecting boundary conditions\cite{Maggio:2021uge}, as were discussed in Sec.~\ref{NHBCsec}.  
In that case we see from \eqref{scattnh}, aside from the possible resonant structure arising from the denominator, that $|\Scatt_{l}(n\omega_0)|\propto |T_{n \omega_0,l}|^2$ gives the leading behavior, resulting in a frequency-independent $|\ccqo| \sim |\tilde{\calr}|$ from \eqref{Ccqo}.  
Possible resonant features arising from the subleading corrections in \eqref{scattnh}, \eqref{NHBCexp} have been argued to
 be too sharp to be observable \cite{Cardoso:2019nis, Maggio:2021uge, Cardoso:2022fbq}. Then, the bounds  $|\tilde{\calr}| \lesssim 10^{-4}$ from \cite{Maggio:2021uge}, also found in the case of spin $\chi=0.8$, compare directly with those of \cite{Datta:2019epe}.\footnote{To compare results, it is important to be careful with definition of the physical quantities.  We have shown that $\ccqo$ is to leading order {\it linear} in the reflection coefficient $\tilde{\cal R}$, as seen from \eqref{scattnh} and \eqref{Cscal}.  Ref.~\cite{Datta:2019epe} instead assumed that $\ccqo$ was quadratic in the reflection coefficient.  Comparison in terms of $\ccqo$ thus yields approximate agreement, modulo the detailed structure of possible resonances arising from the denominator of  \eqref{scattnh}.}  Ref.~\cite{Sago:2021iku} likewise investigated models with near-horizon reflection, giving similar results for sensitivity to the reflection coefficient.

One can ask the question of what range of orbits, if any, dominates the sensitivity.  While higher-frequency, small $r_0$ orbits will be more sensitive to the near-horizon physics, fewer such orbits will contribute to the signal. In the constant $\ccqo$ examples, the interplay between both effects ensures that no one specific frequency contributes particularly strongly to the bound. However, a significant drop occurs around the innermost stable circular orbit, where the binding energy reaches a minimum. 

In more general models,  one expects  $\ccqo$ to be frequency dependent, and so such considerations are model dependent.     They also depend on the astrophysics.  Aside from the dependence of frequencies that are probed on the spin\cite{Babak:2017tow},  it has been proposed that binary systems with mass-ratio $10^{8}$ could be detectable in the LISA band \cite{Amaro-Seoane:2019umn}. These would hardly evolve over years of LISA data. Therefore, they would yield bounds on $\ccqo$ without needing extra theoretical input on the frequency dependence. On the other hand, this would limit the parameter space that can be tested.

We would like to use  sensitivity to $\ccqo$ to investigate possible values of the scattering parameters $\Scatt_{AA'}(\omega, \omega')$ representing the deviations from classical BH behavior, 
and ultimately constrain the parameters of the underlying interactions responsible for such deviations.  Prior to investigating models for such physics, we can give a preliminary analysis of the possible sensitivity to such interactions via this approach.
 
While it is not possible in full generality to invert the bounds on $\ccqo$ for the individual parameters $\Scatt_{AA'}(\omega, \omega')$, we can begin to better understand constraints by considering the case where a single mode provides the dominant contribution.  Focusing on the simpler scalar case, which we have argued provides a useful proxy with much of the same structure as the gravitational wave case, consider the situation where a single mode, say with values $l_*$, $m_*$, is dominant.  
If this is the case, let us introduce the relative contribution of that single mode to the black hole horizon flux $\cE_{l_* m_*}(\omega_0)$
\begin{equation}
\cE_{l_* m_*} = \frac{|Y_{l_* m_*}(\pi/2,0)|^2  |u_{m_*\omega_0,l_*}^{up\,}(r_0)|^2 |T_{m_*\omega_0,l_*}|^2}{\sum_{ln}|Y_{ln}(\pi/2,0)|^2 |u_{n\omega_0,l}^{up}(r_0)|^2 |T_{n\omega_0,l}|^2  }	\, .
\end{equation}
Then the relation between the dominant $\Scatt_{l_*}(\omega)$ and $|\ccqo|$ is  
\begin{equation}\label{eqn:domC}
|\ccqo| \sim \frac{\cE_{l_* m_*}}{|T_{m_*\omega_0,l_*}|^2} |\Scatt_{l_*}(m_*\omega_0)|  \, ,
\end{equation} 
where instead of the contribution from a relative phase,  which is likely highly model dependent, we have used the root mean square over such phases. 

To give an example, at around $r_0 = 10M$, using\footnote{In fact, this is only true if one accounts for both $m=\pm 2$ which contribute equally. One finds $\cE_{2 2}+ \cE_{2 -2} \approx 0.9$.} $\cE_{2 2} \sim 1$ and $|T_{2 \omega_0,2}|^2 \sim 10^{-7}$ (values for the gravitational case), one observes that the conservative percent level bounds on $\ccqo$, as argued for from the phase shift above, would lead to sensitivity to $|\Scatt_{2}| \lesssim 10^{-9}$. The key observation here is that, due to the small transmission factor, even rather weak bounds on $\ccqo$ translate into very tight constraints on $|\Scatt_{l}|$. 

\begin{figure}[t!]
	\begin{center}
		\includegraphics[width=0.48\textwidth]{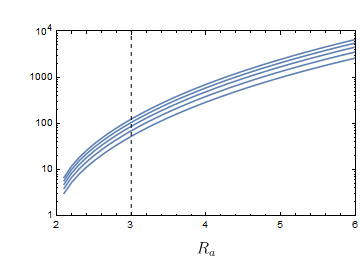} 
		\includegraphics[width=0.48\textwidth]{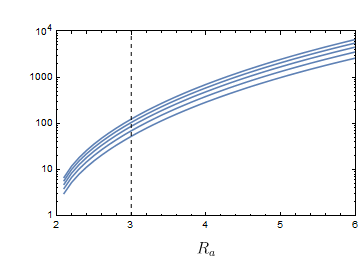} 
		\caption{The ``gain factor'' of the halo model compared to the near-horizon boundary condition $[u_{\omega l}^{\rm in}(R_a)/ u_{\omega l}^{\rm up}(R_a)]/|T_{\omega,l}|^2$ as a function of the halo scale $R_a$ for a range of orbital radii (bottom to top: $6M$, $7M$, $8M$, $9M$, $10M$). On the left for the $l=m=2$-mode while on the right for the $l=2$, $m=1$ mode. }
		\label{fig:gain}
	\end{center}
\end{figure}

We can also investigate sensitivity to models for scattering from a halo of a BH, which we argued in Sec.~\ref{scatta} may more realistically represent the range of quantum interactions responsible for unitary evolution of quantum black holes.  In the simple halo model of Sec.~\ref{scatta}, with $\ccqo$ given by combining \eqref{Ccqo} and \eqref{refdef}, one likewise  finds that the influence of  the scattering parameters ${\cR}_{\omega l}(R_a)$ on $\Scatt_l(\omega)$ is relatively {\it enhanced} by the absence of a full suppression by $|T_{\omega,l}|^2$; this is understood as due to the scattering wave having to tunnel less far before interacting, and the resulting scattered wave also correspondingly having a smaller tunneling suppression.  This 
``gain factor'' can be seen from \eqref{refdef} in terms of  the relative ratio $[u_{\omega l}^{\rm in}(R_a)/ u_{\omega l}^{\rm up}(R_a)]/|T_{\omega,l}|^2$. For  orbits ranging from $r_0 = 6M$ to $r_0 = 10M$ and a range of $R_a$ we present this ratio in figure \ref{fig:gain}. Notice also that the difference is more pronounced for lower frequencies.

As a representative example one can take $R_a$ to be near the photon ring. In this case, for given interaction strength, $\Scatt_l$ is well over an order of magnitude bigger in the halo model compared to a near-horizon boundary model with an equal reflectivity parameter. To translate this to $\ccqo$, let us assume a constant $\cR(R_a)$ in \eqref{refdef} and set $R_a = 3M$. For frequencies associated to $r_0 \approx 10M$, one finds, using \eqref{Ccqo}, \eqref{refdef}, and again numerically evaluating black hole quantities using the Black Hole Toolkit, 
 $|\ccqo| \sim 10^2 |{\cR}(R_a)|$. Closer to $r_0 \approx 6M$, this may be about a factor of two less, as illustrated in figure \ref{fig:gain}. Nevertheless, integrating \eqref{eqn:phaseshift} between these ranges, we find $2 \Delta \phi_0 \sim   10^{-1} \frac{M}{\mu} |{\cR}(R_a)|$, so for an EMRI with central Schwarzschild BH, this would correspond to a sensitivity
down to $|{\cR}(R_a)| \lesssim 10^{-4}$. If one compares to the case of a rotating BH with spin $\chi\sim0.8$, as discussed above, the preceding discussion and bounds of \cite{Datta:2019epe,Maggio:2021uge} suggest an $\calo(10^2)$ enhancement of sensitivity, which would give a sensitivity down to the range $|{\cR}(R_a)| \lesssim 10^{-6}$.
These are of course preliminary estimates,
which should be further explored in study of more complete models for interactions, and more thorough treatment of their gravitational wave signals.   It is also important to better understand the extent to which such sensitivity can be achieved when faced with more realistic data and astrophysics.

Eccentric and inclined orbits are one aspect of  more general situations, resulting in a richer frequency content \cite{Hughes:2021exa,Drasco:2005kz,Speri:2023jte}. Based on the behavior of the transmission factors $T_{\omega l}$, the resulting additional power in higher frequency modes suggests one might expect to find even stronger constraints for scattering effects from quantum black holes in these cases.\footnote{As this paper was being finalized, we received \cite{Datta:2024vll}, studying eccentric orbits and describing roughly similar constraints to \cite{Datta:2019epe}.}

\section{Conclusion and outlook}

We have described a general parameterization of the physical effects of deviations of a compact quantum object from a classical black hole, in  terms of the scattering properties of the object.  This extends and generalizes study of specific models for modifications of classical black hole behavior.  Such scattering is in principle observable in idealized circumstances, but is not directly observable with current experimental circumstances.  However, this description of the modification to scattering can then be related to an effect on the gravitational wave signal from a binary of such objects.  In particular, we have found that the effect on the signal deviation is linear in the scattering deviation.  This, together with the ``amplifier" of many orbital cycles during inspiral, indicates a potential sensitivity to small deviations that we have begun to quantify in this paper.  In doing so, we have made connection to related results in specific models, beginning with the work of \cite{Datta:2019epe}.

The current work leaves a number of projects for future research.

First, much of the analysis of this paper was carried out for the simplifying case of scalar radiation on a static, spherically symmetric background.  However, as seen in Section \ref{sec:grav}, gravitational radiation for a CQO with angular momentum does not present major conceptual differences; the setup is entirely analogous.  That still leaves work to be done, specifically connecting our treatment more directly to the Teukolsky formalism, and to numerical methods for studying gravitational wave signals, including for the case of objects of nonzero spin.  Moreover, treatment of inclined and eccentric orbits is also needed to compare with typical astrophysical examples.  Such treatment should then, in turn, connect to a more detailed analysis of possible observational constraints from inspirals expected to be observed by LISA and other future detectors.

This paper has also focused on the simplifying EMRI limit, and it is important to translate this treatment to the the comparable mass binaries relevant for LIGO. The post-Newtonian and post-Minkowskian modeling of the early binary inspiral should be well-suited for this \cite{Goldberger:2004jt,Foffa:2011ub,Porto:2016pyg,Bern:2019nnu,Foffa:2019yfl,Bern:2019crd,Bern:2020buy,Kalin:2020mvi,Kalin:2020fhe,Bern:2020uwk,Goldberger:2020fot,Liu:2021zxr,Dlapa:2021npj,Cho:2021arx,Bern:2021yeh,Cho:2022syn,Bern:2022kto,Bern:2022jvn,Dlapa:2022lmu,Dlapa:2023hsl,Chia:2024bwc} and could be of direct use after relating $\Scatt_{l}(\omega)$ in the limit $\omega M \ll 1$ to the adiabatic tidal deformabilities entering into those approaches.  A proposed approach is to parameterize the
absorption into both bodies separately, and take into account the implied change in the binary binding energy, and resulting corrections to the gravitational wave signal.  Going beyond this to a treatment of merger seems more difficult, though  perhaps progress near the merger can be made with a suitable extension of the effective one-body (EOB) formalism \cite{Buonanno:1998gg,Damour:2001tu,Buonanno:2007pf,Pan:2011gk,Pan:2013rra}.  Modifications to the quasinormal modes that govern the ringdown may also be encoded in the poles of $\Scatt_{l}(\omega)$.

Finally, going beyond this work, it is important to understand better the connection of detailed physical models for quantum black holes (or other modifications to classical black holes) to the scattering amplitudes parameterized in this paper.  Such a relation has been illustrated in the simplified model of scattering from a potential barrier near the black hole, but such models are expected to be oversimplified.  A similar analysis could be performed for more complicated models in the literature now or arising in future work.  In particular, it has been argued\cite{Giddings:2017mym} that interactions needed to unitarize black hole evolution can also have an $\calo(1)$ effect on propagating gravitons near a black hole.  An important topic for future work is additional characterization of these interactions, and relating these to the resulting modifications to scattering amplitudes, as for example parameterized in this paper.


\section*{Acknowledgments}
This material is based upon work supported in part by the Heising-Simons Foundation under grant \#2021-2819, and by the U.S. Department of Energy, Office of Science, under Award Number {DE-SC}0011702. This work makes use of the Black Hole Perturbation Toolkit.  We thank Y. Chen, Scott Hughes, and B. Seymour for useful conversations.

\appendix

\newpage


\section{Black hole Green function and energy loss}\label{app:BHGF}

In this appendix we review the calculation of the energy emitted by a source to the scalar wave equation, \eqref{eqn:KG} in a black hole background.  This can be calculated by using a Green function to derive the scalar solution to \eqref{eqn:KG}, and calculating the resulting energy-momentum tensor and fluxes.

The scalar Green function satisfies the defining equation \eqref{GFeq} and gives  the solution \eqref{PhiJ}, where the background metric and boundary conditions are those of the BH.  The Green function can be calculated by a matching procedure.  If $x=(t,r,\theta,\phi)$, then for either $r>r'$ or $r<r'$ it is a solution of the homogeneous equation \eqref{eqn:KG} with $J=0$.  This means it can be expanded in terms of homogeneous solutions \eqref{phidef} of the Regge-Wheeler equation \eqref{eqn:RW} in the two regions,
\beq
G^{bh}\left(x,x'\right) = \begin{cases} 
G^{<} = \sum_{lm}\int_0^\infty d\omega K^<_{\omega lm}(x') \phi^<_{\omega l m}(x) + cc  \qquad &r < r' \\
G^{>} = \sum_{lm}\int_0^\infty d\omega K^>_{\omega lm}(x') \phi^>_{\omega l m}(x) + cc  \qquad & r > r'
\end{cases} \, .
\eeq
Boundary conditions are enforced by requiring a pure outgoing solution at infinity, and pure ingoing at the horizon, which correspond to
\beq
\phi^<_{\omega l m}(x) = \phi^{in}_{\omega l m}(x)\quad ,\quad  \phi^>_{\omega l m}(x) =  \phi^{up}_{\omega l m}(x) \ .
\eeq

Continuity in $G$ at $r=r'$ implies
\beq
\sum_{lm}\int_0^\infty d\omega K^<_{\omega lm}(r,t',\theta',\phi') \phi^{in}_{\omega l m}(x)+ cc  = \sum_{lm}\int_0^\infty d\omega K^>_{\omega lm}(r,t',\theta',\phi') \phi^{up}_{\omega l m}(x)+ cc\ 
\eeq
or
\beq\label{GF1}
K^<_{\omega lm}(r,t',\theta',\phi') u^{in}_{\omega l}(r) = K^>_{\omega lm}(r,t',\theta',\phi') u^{up}_{\omega l}(r)\ ,
\eeq
using \eqref{phidef}.
A second equation follows from integrating \eqref{GFeq} across $r=r'$, which gives
\beq
\left(1-\frac{R}{r}\right) r^2\left[\partial_r G^>(x,x') - \partial_r G^<(x,x') \right]_{r=r'} = \delta(t-t') \frac{\delta(\theta-\theta') \delta(\phi-\phi')}{\sin \theta}\ ,
\eeq
Integrating over $t, \Omega$ against $e^{i\omega t}Y_{lm}^*(\theta,\phi)$ and using the normalization 
\beq
\int d\Omega Y^*_{lm} Y_{l'm'} = \delta_{ll'}\delta_{mm'}
\eeq
then yields at $r=r'$
\beq \label{GF2}
K^>_{\omega lm}(x')\partial_{r_*} u^{up}_{\omega l} - K^<_{\omega lm}(x')\partial_{r_*}u^{in}_{\omega l} = \frac{e^{i\omega t'}}{2\pi r} Y_{lm}^*(\Omega')\ .
\eeq
Equations \eqref{GF1} and \eqref{GF2} imply that 
\beq
K_{\omega lm}(x')= \frac{k_{\omega l}(r')}{2\pi r} e^{i\omega t'} Y_{lm}^*(\Omega')\ ,
\eeq
with
\beq \label{GF3}
k^>_{\omega l} u^{up}_{\omega l}=k^<_{\omega l} u^{in}_{\omega l} \quad ; \quad   k_{\omega l}^>\partial_{r_*} u^{up}_{\omega l} -  k_{\omega l}^<\partial_{r_*} u^{in}_{\omega l} = 1\ .
\eeq
Equations  \eqref{GF3} are then solved by
\beq
k^>_{\omega l}= \frac{u^{in}_{\omega l}}{W_{r_*}[u^{in}_{\omega l}, u^{up}_{\omega l}]}\quad ,\quad k^<_{\omega l}= \frac{u^{up}_{\omega l}}{W_{r_*}[u^{in}_{\omega l}, u^{up}_{\omega l}]}\ .
\eeq
Here the Wronskian 
\beq
W_{r_*}[u^{in}_{\omega l},u^{up}_{\omega l}] = u^{in}_{\omega l} \partial_{r*} u^{up}_{\omega l} - \partial_{r_*} u^{in}_{\omega l} u^{up}_{\omega l} = 2i\omega
\eeq 
is a constant by the Regge-Wheeler equation \eqref{eqn:RW}, and thus may be evaluated using the asymptotic solutions at $r_*=\infty$ with the normalization conventions given in \eqref{eqn:Rupdowndefsscal} and \eqref{ininfty}.  We then find
\beq
K^>_{\omega lm} = \frac{1}{4\pi i \omega} \phi_{\omega lm}^{out*}\quad ,\quad K^<_{\omega lm} = \frac{1}{4\pi i \omega} \phi_{\omega lm}^{down*}\ ,
\eeq
where we have used $u^{out*}_{\omega l}(r) = u^{in}_{\omega l}(r)$ and $u^{down*}_{\omega l}(r) = u^{up}_{\omega l}(r)$ from \eqref{ccreln}.

The result is the BH Green function
\beq
G^{bh}(x,x') = 
\begin{cases} 
G^{<} = \sum_{lm}\int_0^\infty \frac{d\omega}{4\pi i \omega} \phi^{down*}_{\omega lm}(x') \phi^{in}_{\omega l m}(x) + cc  \qquad &r < r' \\
G^{>} = \sum_{lm}\int_0^\infty \frac{d\omega}{4\pi i \omega}  \phi^{out*}_{\omega lm}(x') \phi^{up}_{\omega l m}(x) + cc  \qquad & r > r'
\end{cases} \, .
\eeq
This gives a solution to the inhomogeneous equation \eqref{eqn:KG} in the BH background of the form
\beq\label{BHphi}
\Phi^{bh}_J(x) = \int dV_4' G^{bh}(x,x') J(x') = \begin{cases} 
\sum_{lm}\int_0^\infty d\omega Z^{down}_{\omega l m}[J] \phi^{in}_\olm(x) + cc  \qquad &r < r' \\
\sum_{lm}\int_0^\infty d\omega Z^{out}_\olm[J] \phi^{up}_\olm(x) + cc \qquad & r > r' \end{cases} \, 
\eeq
where $Z^{down}_\olm[J]$ was defined in \eqref{Zdefs}, and $Z^{out}_\olm[J]$ in \eqref{BHcont}.

Given the source $J(x)$, say corresponding to an orbiting body, we would like to calculate the radiated energy carried by the field $\Phi^{bh}_J$.  As in the main text, we assume that the source $J$ has support within a range of radii $r_1<r<r_2$.  The outward energy flux through a sphere of radius $r$ is 
\begin{equation}\label{eqn:dEflux}
		 -\int_{S_2} r^2 d\Omega \, \,  T^{r}{}_t \, ,
\end{equation}
with stress tensor \eqref{stressT}, 
and we have such contributions at radii above and below the source, decreasing the energy $E$ in the source region.  Consider the contribution at radius $r\geq r_2$.  Inserting the lower line of \eqref{BHphi} into $T_{\mu\nu}$ and time 
averaging over long time $T$ gives the radiated power
\beq
-\Big\langle\frac{d E}{d t}(r)\Big\rangle_> = i \sum_{lm} \int_0^\infty d\omega d\omega' Z^{out}_\olm[J] Z^{out*}_{\omega' lm}[J] r\left[\omega u_{\omega l}^{up}  \partial_{r_*} \left(\frac{u^{up *}_{\omega' l}}{r}\right) - \omega' \partial_{r_*}\left( \frac{u_{\omega l}^{up}}{r}\right) u^{up *}_{\omega' l}\right] \Delta_T(\omega -\omega')
\eeq
where we drop contributions that average to zero in the long time  limit, and the angular average matches angular quantum numbers.  We have also used the definition \eqref{Deltadef}, which gives in the long-time limit $\Delta_T(\omega-\omega')\propto \delta(\omega-\omega')$, as discussed in Sec. \ref{sec:energyloss} and Appendix \ref{app:averaging}.  This results in
\beq\label{dEdtG}
-\Big\langle\frac{d E}{d t}(r)\Big\rangle_> = i \sum_{lm} \int_0^\infty d\omega d\omega' \Delta_T(\omega-\omega') \big |Z^{out}_\olm[J]\big|^2 \omega W_{r_*}(u^{up}_{\omega l}, u^{up*}_{\omega l})\ .
\eeq
The Wronskian is
\beq
W_{r_*}(u^{up}_{\omega l},u^{up*}_{\omega l})= -2i\omega\ .
\eeq
Correspondingly, we have 
\beq
-\int_{S^2} d\Omega\, T^r_t(\phi^{up}_{\omega lm},\phi^{up*}_{\omega lm}) = \omega^2\quad ,\quad -\int_{S^2} d\Omega\, T^r_t(\phi^{down}_{\omega lm},\phi^{down*}_{\omega lm}) = -\omega^2\ ,
\eeq
which are $r$-independent due to conservation.  Eq.~\eqref{dEdtG} becomes 
\beq
-\Big\langle\frac{d E}{d t}(r)\Big\rangle_> =  \sum_{lm} \int_0^\infty d\omega d\omega' \Delta_T(\omega-\omega') 2 \omega^2\big |Z^{out}_\olm[J]\big|^2 ,
\eeq
which can be interpreted by saying that each mode contributes $2\omega^2$ to $-dE/dt$ (in our normalization), and the $Z_\olm$ give the amplitudes for excitation of individual modes. 

An analogous calculation for $r<r_1$,  with a sign change to account for energy escaping the source region, likewise gives
\beq\label{EBHhor}
-\Big\langle\frac{d E}{d t}(r)\Big\rangle_ < = \sum_{lm} \int_0^\infty d\omega d\omega' \Delta_T(\omega-\omega')2\omega^2 \big |Z^{down}_\olm[J]\big|^2 |T_{\omega l}|^2 \ =-\Big\langle\frac{d E}{d t}\Big\rangle_ {BH,hor}
\eeq
where the transmission coefficient $T_{\omega l}$ enters the Wronskian through the normalization \eqref{intohor},
resulting in a total average power emitted
\beq\label{EBHtot}
-\Big\langle\frac{d E}{d t}\Big\rangle_{BH,tot} = \sum_{lm} \int_0^\infty d\omega d\omega' \Delta_T(\omega-\omega') 2\omega^2 \left\{ \big |Z^{out}_\olm[J]\big|^2 + |T_{\omega l}|^2 \big |Z^{down}_\olm[J]\big|^2\right\}\ .
\eeq
In the case of a periodic orbit, as described in Section~\ref{genframe}, this becomes
\beq\label{Eemper}
-\Big\langle\frac{d E}{d t}\Big\rangle = \sum_{nlm} 2 \omega_n^2  \left\{ \big |Z^{out}_{nlm}[J]\big|^2 + |T_{\omega_n l}|^2 \big |Z^{down}_{nlm}[J]\big|^2\right\}\ .
\eeq

\begin{figure}[t!]
	\begin{center}
		\includegraphics[width=0.68\textwidth]{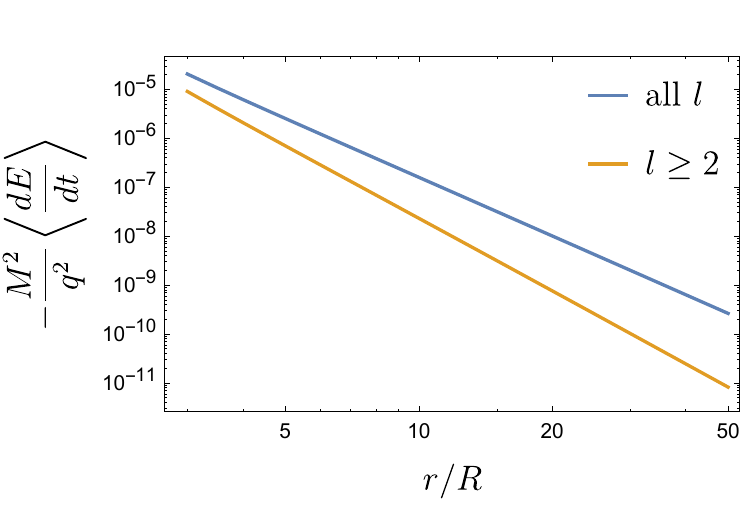} 
		\caption{Energy lost to a scalar particle of charge $q$ circularly orbiting a black hole at radius $r$, both including $l<2$ contributions (blue) and ignoring them (orange).}
		\label{fig:Efluxscalar}
	\end{center}
\end{figure}

In comparing the scalar radiated energy to that of metric perturbations,  it may be helpful to consider the relative contribution of the $l\geq2$ modes to the total.  
In Figure \ref{fig:Efluxscalar} we present as an example the energy lost to the scalar wave emission from circular orbits as a function of radius (up to the overall ratio $q^2/M^2$, with $q$ the scalar charge), with and without the $l<2$ contributions.

 \section{Adiabatic or two-timescale time averages}\label{app:averaging}
 
  In this appendix, some further comments are made on the time-averaging procedure in the main text. On timescales small compared to the radiation reaction timescale $T_{rad} \sim M^2/\mu$, with $M$ the mass of the supermassive object and $\mu$ the mass of the stellar mass companion, the source associated to an extreme mass-ratio black hole binary would be (multiply) periodic; with periods related to the radial, longitudinal, and  azimuthal motion. The frequency content is thus effectively  discretized, as described in the main text. A caveat is that $\omega_n$ is in fact spread out over a frequency range $\sim 1/T_{rad}$; only for a test-mass is the orbital frequency content truly discrete. 
 
 In the periodic case, one naturally averages over the period
 \begin{equation}\label{eqn:periodicaverage}
 \left\langle e^{-i \left(\omega_n-\omega_{n'}\right) t}\right\rangle_{period} = \frac{1}{T} \int_0^{T} e^{-i \left(\omega_n-\omega_{n'}\right) t} = \delta_{n n'} \, ,
 \end{equation}
 where $\omega_1 = 2\pi/T$ represents the fundamental frequency. \eqref{eqn:periodicaverage} still holds if we average over $NT$ for arbitrary positive integer $N$. If there are multiple but commensurate fundamental frequencies, one would use the smallest common period. If the fundamental frequencies are not commensurate or a continuum of frequencies is relevant for other reasons, say by mixing due to scattering as in the main text, one can take a limit where $N \to \infty$; let $\omega_n \neq \omega$ then
 \begin{equation}
 \left\langle e^{-i \left(\omega_n-\omega\right) t}\right\rangle_{N \, period}  = \lim_{N \to \infty} \frac{1}{N T} \int_0^{N T} e^{-i \left(\omega_n-\omega\right) t} \sim  \frac{1}{N T} \ll 1\, .
 \end{equation} 
 However, as mentioned above, even for exactly discrete test-mass orbital frequencies, the radiation reaction timescale $T_{rad}$ provides a natural physical cutoff, suggesting instead
 \begin{equation}\label{eqn:sincaverage}
 \left\langle  e^{i (\omega'-\omega) t} \right\rangle_{rad} = \frac{1}{T_{rad}} \int^{T_{rad}/2}_{-T_{rad}/2} dt \,  e^{i (\omega'-\omega) t} = \, \text{sinc}\left\lbrack\left(\omega'-\omega\right) \frac{T_{rad}}{2}\right\rbrack \, .
 \end{equation}
 with
 \begin{equation}
 \text{sinc}(x) = \frac{\sin{x}}{x} \, , \qquad \int_{-\infty}^{\infty} dx \, \text{sinc}(x) = \pi \, .
 \end{equation}
 This has the well-known limit \cite{NIST:DLMF}
 \begin{equation}\label{eqn:sinclimit}
 \lim_{T_{rad} \to \infty}  \frac{T_{rad}}{2} \text{sinc}\left\lbrack\left(\omega'-\omega\right) \frac{T_{rad}}{2}\right\rbrack = \pi \delta(\omega'-\omega) \, .
 \end{equation}
 Therefore, as $T_{rad} \to \infty$ or more precisely $(\omega'-\omega) T_{rad} \to \infty$ , one finds
 \begin{equation}\label{eqn:sincdelta}
 \left\langle  e^{i (\omega'-\omega) t} \right\rangle_{rad} =  \, \text{sinc}\left\lbrack\left(\omega'-\omega\right) \frac{T_{rad}}{2}\right\rbrack  \to \frac{2\pi}{T_{rad}} \delta(\omega'-\omega) \, .
 \end{equation}
 On the other hand, one ideally chooses an averaging timescale $T_{av}$ intermediate between the characteristic dynamical timescale, and the radiation reaction scale $T_{rad}$
 \begin{equation}\label{eqn:Thierarchy}
 T_{dyn} \ll T_{av} \ll T_{rad} \, .
 \end{equation}
 Consequently, in the main text, we use a time-averaging
 \begin{equation}
 \left\langle f(t) \right \rangle = \frac{1}{T_{av}} \int^{T_{av}/2}_{-T_{av}/2} dt \, f(t) \, .
 \end{equation}
 For a circular orbit $T_{dyn} \sim r_0^{3/2}/\sqrt{M}$, this can be done without much ambiguity. Therefore, we have in fact kept the choice implicit by introducing
 \begin{equation}\label{eqn:Deltadef}
 \Delta_{T}(\omega_1-\omega_2) = \frac{1}{T} \int^{T/2}_{-T/2} dt \,  e^{-i (\omega_1-\omega_2) t}  \, .
 \end{equation}
 Nevertheless, key features are the limit \eqref{eqn:sincdelta} and \eqref{eqn:periodicaverage}; it behaves as a $\delta$-function for a continuous spectrum as $T_{rad}(\omega'-\omega) \gg T_{av}(\omega'-\omega) \to \infty$, and it recovers the periodic result. For the latter note
 the ``discrete'' frequencies are considered ``equal'' only if $\omega_{n'}-\omega_n \lesssim \frac{2 \pi}{T_{rad}}$, then
 \begin{equation}
 \text{sinc}\left\lbrack\left(\omega_{n'}-\omega_n\right)\frac{T_{av}}{2}\right\rbrack \sim 1 + \calo\left(\left( \frac{T_{av}}{T_{rad}}\right)^2\right) \, ,
 \end{equation} 
 If they are not equal and not resonant in any way, one should instead expect $\omega_{n'}-\omega_n \sim \frac{2 \pi}{T_{dyn}}$ or
 \begin{equation}
 \text{sinc}\left\lbrack\left(\omega_{n'}-\omega_n\right)\frac{T_{av}}{2}\right\rbrack \sim \calo\left( \frac{T_{ dyn}}{T_{av}}\right) \, ,
 \end{equation} 
 Therefore, in the discrete case as $T_{rad}(\omega_{n'}-\omega_n) \gg T_{av}(\omega_{n'}-\omega_n) \to \infty$,
 \begin{equation}
 \text{sinc}\left\lbrack\left(\omega_{n'}-\omega_n\right)\frac{T_{av}}{2}\right\rbrack \to \delta_{n' n} \, ,
 \end{equation}
 As desired. While we have thus shown that the choice \eqref{eqn:Deltadef} for $\Delta_{T}$ is a good one, the result in the main text does not depend on the details of this choice.
 
 Let us conclude by pointing out that, when additional orbital frequencies are involved, subtleties arise related to orbital resonances. In essence, a low beating frequency develops between otherwise high-frequency orbital modes, leading to a breakdown of the hierarchy of scales \eqref{eqn:Thierarchy}. The implications of such resonances for gravitational wave astronomy are under active investigation \cite{Flanagan:2012kg,Brink:2013nna,Nasipak:2021qfu,Destounis:2021rko,Speri:2021psr,Nasipak:2022xjh,Pan:2023wau}. 
 
\providecommand{\href}[2]{#2}\begingroup\raggedright\endgroup

\bibliographystyle{toine}

\begin{thebibliography}{100}
	
	\bibitem{Hawk}
	S.~W. Hawking, \emph{{Particle Creation by Black Holes}}, Commun. Math. Phys.
	{\bf 43} (1975) 199--220,
	[Erratum: Commun.Math.Phys. 46, 206 (1976)]
	
	\bibitem{Hawkunc}
	S.~W. Hawking, \emph{{Breakdown of Predictability in Gravitational Collapse}},
	Phys. Rev. D {\bf 14} (1976)
	2460--2473
	
	\bibitem{EventHorizonTelescope:2019dse}
	{\bf Event Horizon Telescope} Collaboration, K.~Akiyama {\em et al.},
	\emph{{First M87 Event Horizon Telescope Results. I. The Shadow of the
			Supermassive Black Hole}}, Astrophys. J. Lett. {\bf 875} (2019) L1,
	\href{http://www.arXiv.org/abs/1906.11238}{{\tt 1906.11238}}
	
	\bibitem{EventHorizonTelescope:2022wkp}
	{\bf Event Horizon Telescope} Collaboration, K.~Akiyama {\em et al.},
	\emph{{First Sagittarius A* Event Horizon Telescope Results. I. The Shadow of
			the Supermassive Black Hole in the Center of the Milky Way}}, Astrophys. J.
	Lett. {\bf 930} (2022), no.~2,
	L12
	
	\bibitem{LIGOScientific:2018mvr}
	{\bf LIGO Scientific, Virgo} Collaboration, B.~P. Abbott {\em et al.},
	\emph{{GWTC-1: A Gravitational-Wave Transient Catalog of Compact Binary
			Mergers Observed by LIGO and Virgo during the First and Second Observing
			Runs}}, Phys. Rev. X {\bf 9} (2019), no.~3, 031040,
	\href{http://www.arXiv.org/abs/1811.12907}{{\tt 1811.12907}}
	
	\bibitem{LIGOScientific:2020ibl}
	{\bf LIGO Scientific, Virgo} Collaboration, R.~Abbott {\em et al.},
	\emph{{GWTC-2: Compact Binary Coalescences Observed by LIGO and Virgo During
			the First Half of the Third Observing Run}}, Phys. Rev. X {\bf 11} (2021)
	021053,
	\href{http://www.arXiv.org/abs/2010.14527}{{\tt 2010.14527}}
	
	\bibitem{LIGOScientific:2021usb}
	{\bf LIGO Scientific, VIRGO} Collaboration, R.~Abbott {\em et al.},
	\emph{{GWTC-2.1: Deep extended catalog of compact binary coalescences
			observed by LIGO and Virgo during the first half of the third observing
			run}}, Phys. Rev. D {\bf 109} (2024), no.~2, 022001,
	\href{http://www.arXiv.org/abs/2108.01045}{{\tt 2108.01045}}
	
	\bibitem{KAGRA:2021vkt}
	{\bf KAGRA, VIRGO, LIGO Scientific} Collaboration, R.~Abbott {\em et al.},
	\emph{{GWTC-3: Compact Binary Coalescences Observed by LIGO and Virgo during
			the Second Part of the Third Observing Run}}, Phys. Rev. X {\bf 13} (2023),
	no.~4, 041039,
	\href{http://www.arXiv.org/abs/2111.03606}{{\tt 2111.03606}}
	
	\bibitem{Colpi:2024xhw}
	M.~Colpi {\em et al.}, \emph{{LISA Definition Study Report}},
	\href{http://www.arXiv.org/abs/2402.07571}{{\tt 2402.07571}}
	
	\bibitem{Mark:2017dnq}
	Z.~Mark, A.~Zimmerman, S.~M. Du  and Y.~Chen, \emph{{A recipe for echoes from
			exotic compact objects}}, Phys. Rev. D {\bf 96} (2017), no.~8, 084002,
	\href{http://www.arXiv.org/abs/1706.06155}{{\tt 1706.06155}}
	
	\bibitem{Maggio:2018ivz}
	E.~Maggio, V.~Cardoso, S.~R. Dolan  and P.~Pani, \emph{{Ergoregion instability
			of exotic compact objects: electromagnetic and gravitational perturbations
			and the role of absorption}}, Phys. Rev. D {\bf 99} (2019), no.~6, 064007,
	\href{http://www.arXiv.org/abs/1807.08840}{{\tt 1807.08840}}
	
	\bibitem{Cardoso:2021ehg}
	V.~Cardoso, C.~F.~B. Macedo, K.-i. Maeda  and H.~Okawa, \emph{{ECO-spotting:
			looking for extremely compact objects with bosonic fields}}, Class. Quant.
	Grav. {\bf 39} (2022), no.~3, 034001,
	\href{http://www.arXiv.org/abs/2112.05750}{{\tt 2112.05750}}
	
	\bibitem{Ryan:2022hku}
	M.~Ryan and D.~Radice, \emph{{Exotic compact objects: The dark white dwarf}},
	Phys. Rev. D {\bf 105} (2022), no.~11, 115034,
	\href{http://www.arXiv.org/abs/2201.05626}{{\tt 2201.05626}}
	
	\bibitem{Cunha:2022gde}
	P.~V.~P. Cunha, C.~Herdeiro, E.~Radu  and N.~Sanchis-Gual, \emph{{Exotic
			Compact Objects and the Fate of the Light-Ring Instability}}, Phys. Rev.
	Lett. {\bf 130} (2023), no.~6, 061401,
	\href{http://www.arXiv.org/abs/2207.13713}{{\tt 2207.13713}}
	
	\bibitem{Cardoso:2019rvt}
	V.~Cardoso and P.~Pani, \emph{{Testing the nature of dark compact objects: a
			status report}}, Living Rev. Rel. {\bf 22} (2019), no.~1, 4,
	\href{http://www.arXiv.org/abs/1904.05363}{{\tt 1904.05363}}
	
	\bibitem{LIGOScientific:2019fpa}
	{\bf LIGO Scientific, Virgo} Collaboration, B.~P. Abbott {\em et al.},
	\emph{{Tests of General Relativity with the Binary Black Hole Signals from
			the LIGO-Virgo Catalog GWTC-1}}, Phys. Rev. D {\bf 100} (2019), no.~10,
	104036,
	\href{http://www.arXiv.org/abs/1903.04467}{{\tt 1903.04467}}
	
	\bibitem{LIGOScientific:2021sio}
	{\bf LIGO Scientific, VIRGO, KAGRA} Collaboration, R.~Abbott {\em et al.},
	\emph{{Tests of General Relativity with GWTC-3}},
	\href{http://www.arXiv.org/abs/2112.06861}{{\tt 2112.06861}}
	
	\bibitem{EventHorizonTelescope:2022xqj}
	{\bf Event Horizon Telescope} Collaboration, K.~Akiyama {\em et al.},
	\emph{{First Sagittarius A* Event Horizon Telescope Results. VI. Testing the
			Black Hole Metric}}, Astrophys. J. Lett. {\bf 930} (2022), no.~2, L17,
	\href{http://www.arXiv.org/abs/2311.09484}{{\tt 2311.09484}}
	
	\bibitem{Abedi:2016hgu}
	J.~Abedi, H.~Dykaar  and N.~Afshordi, \emph{{Echoes from the Abyss: Tentative
			evidence for Planck-scale structure at black hole horizons}}, Phys. Rev. D
	{\bf 96} (2017), no.~8, 082004,
	\href{http://www.arXiv.org/abs/1612.00266}{{\tt 1612.00266}}
	
	\bibitem{Ashton:2016xff}
	G.~Ashton, O.~Birnholtz, M.~Cabero, C.~Capano, T.~Dent, B.~Krishnan, G.~D.
	Meadors, A.~B. Nielsen, A.~Nitz  and J.~Westerweck, \emph{{Comments on:
			''Echoes from the abyss: Evidence for Planck-scale structure at black hole
			horizons''}},
	\href{http://www.arXiv.org/abs/1612.05625}{{\tt 1612.05625}}
	
	\bibitem{Westerweck:2017hus}
	J.~Westerweck, A.~Nielsen, O.~Fischer-Birnholtz, M.~Cabero, C.~Capano, T.~Dent,
	B.~Krishnan, G.~Meadors  and A.~H. Nitz, \emph{{Low significance of evidence
			for black hole echoes in gravitational wave data}}, Phys. Rev. D {\bf 97}
	(2018), no.~12, 124037,
	\href{http://www.arXiv.org/abs/1712.09966}{{\tt 1712.09966}}
	
	\bibitem{Ryan:1997hg}
	F.~D. Ryan, \emph{{Accuracy of estimating the multipole moments of a massive
			body from the gravitational waves of a binary inspiral}}, Phys. Rev. D {\bf
		56} (1997)
	1845--1855
	
	\bibitem{Barack:2006pq}
	L.~Barack and C.~Cutler, \emph{{Using LISA EMRI sources to test off-Kerr
			deviations in the geometry of massive black holes}}, Phys. Rev. D {\bf 75}
	(2007) 042003,
	\href{http://www.arXiv.org/abs/gr-qc/0612029}{{\tt gr-qc/0612029}}
	
	\bibitem{Bena:2020uup}
	I.~Bena and D.~R. Mayerson, \emph{{Black Holes Lessons from Multipole Ratios}},
	JHEP {\bf 03} (2021) 114,
	\href{http://www.arXiv.org/abs/2007.09152}{{\tt 2007.09152}}
	
	\bibitem{Fransen:2022jtw}
	K.~Fransen and D.~R. Mayerson, \emph{{Detecting equatorial symmetry breaking
			with LISA}}, Phys. Rev. D {\bf 106} (2022), no.~6, 064035,
	\href{http://www.arXiv.org/abs/2201.03569}{{\tt 2201.03569}}
	
	\bibitem{Mayerson:2022ekj}
	D.~R. Mayerson, \emph{{Gravitational multipoles in general stationary
			spacetimes}}, SciPost Phys. {\bf 15} (2023), no.~4, 154,
	\href{http://www.arXiv.org/abs/2210.05687}{{\tt 2210.05687}}
	
	\bibitem{Binnington:2009bb}
	T.~Binnington and E.~Poisson, \emph{{Relativistic theory of tidal Love
			numbers}}, Phys. Rev. D {\bf 80} (2009) 084018,
	\href{http://www.arXiv.org/abs/0906.1366}{{\tt 0906.1366}}
	
	\bibitem{Sennett:2017etc}
	N.~Sennett, T.~Hinderer, J.~Steinhoff, A.~Buonanno  and S.~Ossokine,
	\emph{{Distinguishing Boson Stars from Black Holes and Neutron Stars from
			Tidal Interactions in Inspiraling Binary Systems}}, Phys. Rev. D {\bf 96}
	(2017), no.~2, 024002,
	\href{http://www.arXiv.org/abs/1704.08651}{{\tt 1704.08651}}
	
	\bibitem{Cardoso:2017cfl}
	V.~Cardoso, E.~Franzin, A.~Maselli, P.~Pani  and G.~Raposo, \emph{{Testing
			strong-field gravity with tidal Love numbers}}, Phys. Rev. D {\bf 95} (2017),
	no.~8, 084014, \href{http://www.arXiv.org/abs/1701.01116}{{\tt 1701.01116}},
	[Addendum: Phys.Rev.D 95, 089901 (2017)]
	
	\bibitem{Chia:2020yla}
	H.~S. Chia, \emph{{Tidal deformation and dissipation of rotating black holes}},
	Phys. Rev. D {\bf 104} (2021), no.~2, 024013,
	\href{http://www.arXiv.org/abs/2010.07300}{{\tt 2010.07300}}
	
	\bibitem{Nollert:1999ji}
	H.-P. Nollert, \emph{{TOPICAL REVIEW: Quasinormal modes: the characteristic
			`sound' of black holes and neutron stars}}, Class. Quant. Grav. {\bf 16}
	(1999)
	R159--R216
	
	\bibitem{Kokkotas:1999bd}
	K.~D. Kokkotas and B.~G. Schmidt, \emph{{Quasinormal modes of stars and black
			holes}}, Living Rev. Rel. {\bf 2} (1999) 2,
	\href{http://www.arXiv.org/abs/gr-qc/9909058}{{\tt gr-qc/9909058}}
	
	\bibitem{Berti:2018vdi}
	E.~Berti, K.~Yagi, H.~Yang  and N.~Yunes, \emph{{Extreme Gravity Tests with
			Gravitational Waves from Compact Binary Coalescences: (II) Ringdown}}, Gen.
	Rel. Grav. {\bf 50} (2018), no.~5, 49,
	\href{http://www.arXiv.org/abs/1801.03587}{{\tt 1801.03587}}
	
	\bibitem{Cardoso:2019mqo}
	V.~Cardoso, M.~Kimura, A.~Maselli, E.~Berti, C.~F.~B. Macedo  and R.~McManus,
	\emph{{Parametrized black hole quasinormal ringdown: Decoupled equations for
			nonrotating black holes}}, Phys. Rev. D {\bf 99} (2019), no.~10, 104077,
	\href{http://www.arXiv.org/abs/1901.01265}{{\tt 1901.01265}}
	
	\bibitem{Cano:2023jbk}
	P.~A. Cano, K.~Fransen, T.~Hertog  and S.~Maenaut, \emph{{Quasinormal modes of
			rotating black holes in higher-derivative gravity}}, Phys. Rev. D {\bf 108}
	(2023), no.~12, 124032,
	\href{http://www.arXiv.org/abs/2307.07431}{{\tt 2307.07431}}
	
	\bibitem{Cano:2023tmv}
	P.~A. Cano, K.~Fransen, T.~Hertog  and S.~Maenaut, \emph{{Universal Teukolsky
			equations and black hole perturbations in higher-derivative gravity}}, Phys.
	Rev. D {\bf 108} (2023), no.~2, 024040,
	\href{http://www.arXiv.org/abs/2304.02663}{{\tt 2304.02663}}
	
	\bibitem{Maselli:2023khq}
	A.~Maselli, S.~Yi, L.~Pierini, V.~Vellucci, L.~Reali, L.~Gualtieri  and
	E.~Berti, \emph{{Black hole spectroscopy beyond Kerr: Agnostic and
			theory-based tests with next-generation interferometers}}, Phys. Rev. D {\bf
		109} (2024), no.~6, 064060,
	\href{http://www.arXiv.org/abs/2311.14803}{{\tt 2311.14803}}
	
	\bibitem{Cardoso:2017cqb}
	V.~Cardoso and P.~Pani, \emph{{Tests for the existence of black holes through
			gravitational wave echoes}}, Nature Astron. {\bf 1} (2017), no.~9, 586--591,
	\href{http://www.arXiv.org/abs/1709.01525}{{\tt 1709.01525}}
	
	\bibitem{Price:2017cjr}
	R.~H. Price and G.~Khanna, \emph{{Gravitational wave sources: reflections and
			echoes}}, Class. Quant. Grav. {\bf 34} (2017), no.~22, 225005,
	\href{http://www.arXiv.org/abs/1702.04833}{{\tt 1702.04833}}
	
	\bibitem{Volkel:2017kfj}
	S.~H. V\"olkel and K.~D. Kokkotas, \emph{{Ultra Compact Stars: Reconstructing
			the Perturbation Potential}}, Class. Quant. Grav. {\bf 34} (2017), no.~17,
	175015,
	\href{http://www.arXiv.org/abs/1704.07517}{{\tt 1704.07517}}
	
	\bibitem{Volkel:2018czg}
	S.~H. V\"olkel, \emph{{Inverse spectrum problem for quasi-stationary states}},
	J. Phys. Comm. {\bf 2} (2018), no.~2, 025029,
	\href{http://www.arXiv.org/abs/1802.08684}{{\tt 1802.08684}}
	
	\bibitem{Volkel:2019ahb}
	S.~H. V\"olkel, R.~Konoplya  and K.~D. Kokkotas, \emph{{Inverse problem for
			Hawking radiation}}, Phys. Rev. D {\bf 99} (2019), no.~10, 104025,
	\href{http://www.arXiv.org/abs/1902.07611}{{\tt 1902.07611}}
	
	\bibitem{Volkel:2019gpq}
	S.~H. V\"olkel and K.~D. Kokkotas, \emph{{On the Inverse Spectrum Problem of
			Neutron Stars}}, Class. Quant. Grav. {\bf 36} (2019), no.~11, 115002,
	\href{http://www.arXiv.org/abs/1901.11262}{{\tt 1901.11262}}
	
	\bibitem{Baumann:2018vus}
	D.~Baumann, H.~S. Chia  and R.~A. Porto, \emph{{Probing Ultralight Bosons with
			Binary Black Holes}}, Phys. Rev. D {\bf 99} (2019), no.~4, 044001,
	\href{http://www.arXiv.org/abs/1804.03208}{{\tt 1804.03208}}
	
	\bibitem{Baumann:2019ztm}
	D.~Baumann, H.~S. Chia, R.~A. Porto  and J.~Stout, \emph{{Gravitational
			Collider Physics}}, Phys. Rev. D {\bf 101} (2020), no.~8, 083019,
	\href{http://www.arXiv.org/abs/1912.04932}{{\tt 1912.04932}}
	
	\bibitem{Baumann:2021fkf}
	D.~Baumann, G.~Bertone, J.~Stout  and G.~M. Tomaselli, \emph{{Ionization of
			gravitational atoms}}, Phys. Rev. D {\bf 105} (2022), no.~11, 115036,
	\href{http://www.arXiv.org/abs/2112.14777}{{\tt 2112.14777}}
	
	\bibitem{Baumann:2022pkl}
	D.~Baumann, G.~Bertone, J.~Stout  and G.~M. Tomaselli, \emph{{Sharp Signals of
			Boson Clouds in Black Hole Binary Inspirals}}, Phys. Rev. Lett. {\bf 128}
	(2022), no.~22, 221102,
	\href{http://www.arXiv.org/abs/2206.01212}{{\tt 2206.01212}}
	
	\bibitem{Torres:2022fyf}
	T.~Torres, M.~O.~E. Hadj, S.-Q. Hu  and R.~Gregory, \emph{{Regge pole
			description of scattering by dirty black holes}}, Phys. Rev. D {\bf 107}
	(2023), no.~6, 064028,
	\href{http://www.arXiv.org/abs/2211.17147}{{\tt 2211.17147}}
	
	\bibitem{Giddings:2011ks}
	S.~B. Giddings, \emph{{Models for unitary black hole disintegration}}, Phys.
	Rev. D {\bf 85} (2012) 044038,
	\href{http://www.arXiv.org/abs/1108.2015}{{\tt 1108.2015}}
	
	\bibitem{Giddings:2012gc}
	S.~B. Giddings, \emph{{Nonviolent nonlocality}}, Phys. Rev. D {\bf 88} (2013)
	064023,
	\href{http://www.arXiv.org/abs/1211.7070}{{\tt 1211.7070}}
	
	\bibitem{Giddings:2013kcj}
	S.~B. Giddings, \emph{{Nonviolent information transfer from black holes: A
			field theory parametrization}}, Phys. Rev. D {\bf 88} (2013), no.~2, 024018,
	\href{http://www.arXiv.org/abs/1302.2613}{{\tt 1302.2613}}
	
	\bibitem{Giddings:2014nla}
	S.~B. Giddings, \emph{{Modulated Hawking radiation and a nonviolent channel for
			information release}}, Phys. Lett. B {\bf 738} (2014) 92--96,
	\href{http://www.arXiv.org/abs/1401.5804}{{\tt 1401.5804}}
	
	\bibitem{Giddings:2017mym}
	S.~B. Giddings, \emph{{Nonviolent unitarization: basic postulates to soft
			quantum structure of black holes}}, JHEP {\bf 12} (2017) 047,
	\href{http://www.arXiv.org/abs/1701.08765}{{\tt 1701.08765}}
	
	\bibitem{Giddings:2019vvj}
	S.~B. Giddings, \emph{{Black holes in the quantum universe}}, Phil. Trans. Roy.
	Soc. Lond. A {\bf 377} (2019), no.~2161, 20190029,
	\href{http://www.arXiv.org/abs/1905.08807}{{\tt 1905.08807}}
	
	\bibitem{Giddings:2022ipt}
	S.~B. Giddings, \emph{{Comparing models for a unitary black hole S matrix}},
	Phys. Rev. D {\bf 109} (2024), no.~8, 084055,
	\href{http://www.arXiv.org/abs/2212.14551}{{\tt 2212.14551}}
	
	\bibitem{Barausse:2014tra}
	E.~Barausse, V.~Cardoso  and P.~Pani, \emph{{Can environmental effects spoil
			precision gravitational-wave astrophysics?}}, Phys. Rev. D {\bf 89} (2014),
	no.~10, 104059,
	\href{http://www.arXiv.org/abs/1404.7149}{{\tt 1404.7149}}
	
	\bibitem{LIGOScientific:2016ebw}
	{\bf LIGO Scientific, Virgo} Collaboration, B.~P. Abbott {\em et al.},
	\emph{{Effects of waveform model systematics on the interpretation of
			GW150914}}, Class. Quant. Grav. {\bf 34} (2017), no.~10, 104002,
	\href{http://www.arXiv.org/abs/1611.07531}{{\tt 1611.07531}}
	
	\bibitem{Purrer:2019jcp}
	M.~P\"urrer and C.-J. Haster, \emph{{Gravitational waveform accuracy
			requirements for future ground-based detectors}}, Phys. Rev. Res. {\bf 2}
	(2020), no.~2, 023151,
	\href{http://www.arXiv.org/abs/1912.10055}{{\tt 1912.10055}}
	
	\bibitem{Owen:2023mid}
	C.~B. Owen, C.-J. Haster, S.~Perkins, N.~J. Cornish  and N.~Yunes,
	\emph{{Waveform accuracy and systematic uncertainties in current
			gravitational wave observations}}, Phys. Rev. D {\bf 108} (2023), no.~4,
	044018,
	\href{http://www.arXiv.org/abs/2301.11941}{{\tt 2301.11941}}
	
	\bibitem{Datta:2019epe}
	S.~Datta, R.~Brito, S.~Bose, P.~Pani  and S.~A. Hughes, \emph{{Tidal heating as
			a discriminator for horizons in extreme mass ratio inspirals}}, Phys. Rev. D
	{\bf 101} (2020), no.~4, 044004,
	\href{http://www.arXiv.org/abs/1910.07841}{{\tt 1910.07841}}
	
	\bibitem{Maggio:2021uge}
	E.~Maggio, M.~van~de Meent  and P.~Pani, \emph{{Extreme mass-ratio inspirals
			around a spinning horizonless compact object}}, Phys. Rev. D {\bf 104}
	(2021), no.~10, 104026,
	\href{http://www.arXiv.org/abs/2106.07195}{{\tt 2106.07195}}
	
	\bibitem{Sago:2022bbj}
	N.~Sago and T.~Tanaka, \emph{{Efficient search method of anomalous reflection
			by the central object in an extreme mass-ratio inspiral system by future
			space gravitational wave detectors}}, Phys. Rev. D {\bf 106} (2022), no.~2,
	024032,
	\href{http://www.arXiv.org/abs/2202.04249}{{\tt 2202.04249}}
	
	\bibitem{Sorkin:1991bw}
	R.~D. Sorkin, \emph{{The Gravitational electromagnetic Noether operator and the
			second order energy flux}}, Proc. Roy. Soc. Lond. A {\bf 435} (1991)
	635--644
	
	\bibitem{Iyer:1994ys}
	V.~Iyer and R.~M. Wald, \emph{{Some properties of Noether charge and a proposal
			for dynamical black hole entropy}}, Phys. Rev. D {\bf 50} (1994) 846--864,
	\href{http://www.arXiv.org/abs/gr-qc/9403028}{{\tt gr-qc/9403028}}
	
	\bibitem{chandrasekhar1991einstein}
	S.~Chandrasekhar and V.~Ferrari, \emph{The Einstein pseudo-tensor and the flux
		integral for perturbed static space-times}, Proceedings of the Royal Society
	of London. Series A: Mathematical and Physical Sciences {\bf 435} (1991),
	no.~1895,
	645--657
	
	\bibitem{Ferrari:2011rb}
	V.~Ferrari, \emph{{Gravitational waves from perturbed stars}}, Bull. Astron.
	Soc. India {\bf 39} (2011) 203,
	\href{http://www.arXiv.org/abs/1105.1678}{{\tt 1105.1678}}
	
	\bibitem{Misner:1973prb}
	C.~W. Misner, K.~S. Thorne  and J.~A. Wheeler, {\em {Gravitation}}.
	\newblock W. H. Freeman, San Francisco,
	1973
	
	\bibitem{Weinberg:1972kfs}
	S.~Weinberg, {\em {Gravitation and Cosmology}: {Principles and Applications of
			the General Theory of Relativity}}.
	\newblock John Wiley and Sons, New York,
	1972
	
	\bibitem{Teukolsky:1973ha}
	S.~A. Teukolsky, \emph{{Perturbations of a rotating black hole. 1. Fundamental
			equations for gravitational electromagnetic and neutrino field
			perturbations}}, Astrophys. J. {\bf 185} (1973)
	635--647
	
	\bibitem{Merlin:2016boc}
	C.~Merlin, A.~Ori, L.~Barack, A.~Pound  and M.~van~de Meent, \emph{{Completion
			of metric reconstruction for a particle orbiting a Kerr black hole}}, Phys.
	Rev. D {\bf 94} (2016), no.~10, 104066,
	\href{http://www.arXiv.org/abs/1609.01227}{{\tt 1609.01227}}
	
	\bibitem{Toomani:2021jlo}
	V.~Toomani, P.~Zimmerman, A.~Spiers, S.~Hollands, A.~Pound  and S.~R. Green,
	\emph{{New metric reconstruction scheme for gravitational self-force
			calculations}}, Class. Quant. Grav. {\bf 39} (2022), no.~1, 015019,
	\href{http://www.arXiv.org/abs/2108.04273}{{\tt 2108.04273}}
	
	\bibitem{Maggio:2020jml}
	E.~Maggio, L.~Buoninfante, A.~Mazumdar  and P.~Pani, \emph{{How does a dark
			compact object ringdown?}}, Phys. Rev. D {\bf 102} (2020), no.~6, 064053,
	\href{http://www.arXiv.org/abs/2006.14628}{{\tt 2006.14628}}
	
	\bibitem{Sago:2021iku}
	N.~Sago and T.~Tanaka, \emph{{Oscillations in the extreme mass-ratio inspiral
			gravitational wave phase correction as a probe of a reflective boundary of
			the central black hole}}, Phys. Rev. D {\bf 104} (2021), no.~6, 064009,
	\href{http://www.arXiv.org/abs/2106.07123}{{\tt 2106.07123}}
	
	\bibitem{Allen:1997xj}
	G.~Allen, N.~Andersson, K.~D. Kokkotas  and B.~F. Schutz, \emph{{Gravitational
			waves from pulsating stars: Evolving the perturbation equations for a
			relativistic star}}, Phys. Rev. D {\bf 58} (1998) 124012,
	\href{http://www.arXiv.org/abs/gr-qc/9704023}{{\tt gr-qc/9704023}}
	
	\bibitem{Kokkotas:2000up}
	K.~D. Kokkotas and J.~Ruoff, \emph{{Radial oscillations of relativistic
			stars}}, Astron. Astrophys. {\bf 366} (2001) 565,
	\href{http://www.arXiv.org/abs/gr-qc/0011093}{{\tt gr-qc/0011093}}
	
	\bibitem{Buchdahl:1956zz}
	H.~A. Buchdahl, \emph{{Reciprocal static solutions of the equations of the
			gravitational field}}, Austral. J. Phys. {\bf 9} (1956)
	13--18
	
	\bibitem{GKT}
	S.~B. Giddings, S.~Koren  and G.~Trevi\~no, \emph{{Exploring strong-field
			deviations from general relativity via gravitational waves}}, Phys. Rev. D
	{\bf 100} (2019), no.~4, 044005,
	\href{http://www.arXiv.org/abs/1904.04258}{{\tt 1904.04258}}
	
	\bibitem{Cardoso:2016rao}
	V.~Cardoso, E.~Franzin  and P.~Pani, \emph{{Is the gravitational-wave ringdown
			a probe of the event horizon?}}, Phys. Rev. Lett. {\bf 116} (2016), no.~17,
	171101, \href{http://www.arXiv.org/abs/1602.07309}{{\tt 1602.07309}},
	[Erratum: Phys.Rev.Lett. 117, 089902 (2016)]
	
	\bibitem{Cardoso:2016oxy}
	V.~Cardoso, S.~Hopper, C.~F.~B. Macedo, C.~Palenzuela  and P.~Pani,
	\emph{{Gravitational-wave signatures of exotic compact objects and of quantum
			corrections at the horizon scale}}, Phys. Rev. D {\bf 94} (2016), no.~8,
	084031,
	\href{http://www.arXiv.org/abs/1608.08637}{{\tt 1608.08637}}
	
	\bibitem{Lai:1993di}
	D.~Lai, \emph{{Resonant oscillations and tidal heating in coalescing binary
			neutron stars}}, Mon. Not. Roy. Astron. Soc. {\bf 270} (1994) 611,
	\href{http://www.arXiv.org/abs/astro-ph/9404062}{{\tt astro-ph/9404062}}
	
	\bibitem{Lai:1997wh}
	D.~Lai, \emph{{Dynamical tides in rotating binary stars}}, Astrophys. J. {\bf
		490} (1997) 847,
	\href{http://www.arXiv.org/abs/astro-ph/9704132}{{\tt astro-ph/9704132}}
	
	\bibitem{Tsang:2011ad}
	D.~Tsang, J.~S. Read, T.~Hinderer, A.~L. Piro  and R.~Bondarescu,
	\emph{{Resonant Shattering of Neutron Star Crusts}}, Phys. Rev. Lett. {\bf
		108} (2012) 011102,
	\href{http://www.arXiv.org/abs/1110.0467}{{\tt 1110.0467}}
	
	\bibitem{Cardoso:2019nis}
	V.~Cardoso, A.~del Rio  and M.~Kimura, \emph{{Distinguishing black holes from
			horizonless objects through the excitation of resonances during inspiral}},
	Phys. Rev. D {\bf 100} (2019) 084046,
	\href{http://www.arXiv.org/abs/1907.01561}{{\tt 1907.01561}},
	[Erratum: Phys.Rev.D 101, 069902 (2020)]
	
	\bibitem{Cardoso:2022fbq}
	V.~Cardoso and F.~Duque, \emph{{Resonances, black hole mimickers, and the
			greenhouse effect: Consequences for gravitational-wave physics}}, Phys. Rev.
	D {\bf 105} (2022), no.~10, 104023,
	\href{http://www.arXiv.org/abs/2204.05315}{{\tt 2204.05315}}
	
	\bibitem{Oshita:2019sat}
	N.~Oshita, Q.~Wang  and N.~Afshordi, \emph{{On Reflectivity of Quantum Black
			Hole Horizons}}, JCAP {\bf 04} (2020) 016,
	\href{http://www.arXiv.org/abs/1905.00464}{{\tt 1905.00464}}
	
	\bibitem{Wang:2019rcf}
	Q.~Wang, N.~Oshita  and N.~Afshordi, \emph{{Echoes from Quantum Black Holes}},
	Phys. Rev. D {\bf 101} (2020), no.~2, 024031,
	\href{http://www.arXiv.org/abs/1905.00446}{{\tt 1905.00446}}
	
	\bibitem{AMPS}
	A.~Almheiri, D.~Marolf, J.~Polchinski  and J.~Sully, \emph{{Black Holes:
			Complementarity or Firewalls?}}, JHEP {\bf 02} (2013) 062,
	\href{http://www.arXiv.org/abs/1207.3123}{{\tt 1207.3123}}
	
	\bibitem{SGBoltz}
	S.~B. Giddings, \emph{{Hawking radiation, the Stefan\textendash{}Boltzmann law,
			and unitarization}}, Phys. Lett. B {\bf 754} (2016) 39--42,
	\href{http://www.arXiv.org/abs/1511.08221}{{\tt 1511.08221}}
	
	\bibitem{BoPe}
	R.~Bousso and G.~Penington, \emph{{Islands Far Outside the Horizon}},
	\href{http://www.arXiv.org/abs/2312.03078}{{\tt 2312.03078}}
	
	\bibitem{Hughes:2018qxz}
	S.~A. Hughes, \emph{{Bound orbits of a slowly evolving black hole}}, Phys. Rev.
	D {\bf 100} (2019), no.~6, 064001,
	\href{http://www.arXiv.org/abs/1806.09022}{{\tt 1806.09022}}
	
	\bibitem{deCesare:2023rmg}
	M.~de~Cesare and R.~Oliveri, \emph{{Backreaction of scalar waves on black holes
			at low frequencies}}, Phys. Rev. D {\bf 108} (2023), no.~4, 044050,
	\href{http://www.arXiv.org/abs/2305.04970}{{\tt 2305.04970}}
	
	\bibitem{BHPToolkit}
	\emph{{Black Hole Perturbation Toolkit}t.}
	(\href{http://bhptoolkit.org/}{bhptoolkit.org}).
	
	\bibitem{Taracchini:2014zpa}
	A.~Taracchini, A.~Buonanno, G.~Khanna  and S.~A. Hughes, \emph{{Small mass
			plunging into a Kerr black hole: Anatomy of the inspiral-merger-ringdown
			waveforms}}, Phys. Rev. D {\bf 90} (2014), no.~8, 084025,
	\href{http://www.arXiv.org/abs/1404.1819}{{\tt 1404.1819}}
	
	\bibitem{Finn:2000sy}
	L.~S. Finn and K.~S. Thorne, \emph{{Gravitational waves from a compact star in
			a circular, inspiral orbit, in the equatorial plane of a massive, spinning
			black hole, as observed by LISA}}, Phys. Rev. D {\bf 62} (2000) 124021,
	\href{http://www.arXiv.org/abs/gr-qc/0007074}{{\tt gr-qc/0007074}}
	
	\bibitem{Sasaki:1981sx}
	M.~Sasaki and T.~Nakamura, \emph{{Gravitational Radiation From a Kerr Black
			Hole. 1. Formulation and a Method for Numerical Analysis}}, Prog. Theor.
	Phys. {\bf 67} (1982)
	1788
	
	\bibitem{Nasipak:2023kuf}
	Z.~Nasipak, \emph{{Adiabatic gravitational waveform model for compact objects
			undergoing quasicircular inspirals into rotating massive black holes}}, Phys.
	Rev. D {\bf 109} (2024), no.~4, 044020,
	\href{http://www.arXiv.org/abs/2310.19706}{{\tt 2310.19706}}
	
	\bibitem{Lindblom:2008cm}
	L.~Lindblom, B.~J. Owen  and D.~A. Brown, \emph{{Model Waveform Accuracy
			Standards for Gravitational Wave Data Analysis}}, Phys. Rev. D {\bf 78}
	(2008) 124020,
	\href{http://www.arXiv.org/abs/0809.3844}{{\tt 0809.3844}}
	
	\bibitem{Babak:2017tow}
	S.~Babak, J.~Gair, A.~Sesana, E.~Barausse, C.~F. Sopuerta, C.~P.~L. Berry,
	E.~Berti, P.~Amaro-Seoane, A.~Petiteau  and A.~Klein, \emph{{Science with the
			space-based interferometer LISA. V: Extreme mass-ratio inspirals}}, Phys.
	Rev. D {\bf 95} (2017), no.~10, 103012,
	\href{http://www.arXiv.org/abs/1703.09722}{{\tt 1703.09722}}
	
	\bibitem{Amaro-Seoane:2019umn}
	P.~Amaro-Seoane, \emph{{Extremely large mass-ratio inspirals}}, Phys. Rev. D
	{\bf 99} (2019), no.~12, 123025,
	\href{http://www.arXiv.org/abs/1903.10871}{{\tt 1903.10871}}
	
	\bibitem{Hughes:2021exa}
	S.~A. Hughes, N.~Warburton, G.~Khanna, A.~J.~K. Chua  and M.~L. Katz,
	\emph{{Adiabatic waveforms for extreme mass-ratio inspirals via multivoice
			decomposition in time and frequency}}, Phys. Rev. D {\bf 103} (2021), no.~10,
	104014, \href{http://www.arXiv.org/abs/2102.02713}{{\tt 2102.02713}},
	[Erratum: Phys.Rev.D 107, 089901 (2023)]
	
	\bibitem{Drasco:2005kz}
	S.~Drasco and S.~A. Hughes, \emph{{Gravitational wave snapshots of generic
			extreme mass ratio inspirals}}, Phys. Rev. D {\bf 73} (2006), no.~2, 024027,
	\href{http://www.arXiv.org/abs/gr-qc/0509101}{{\tt gr-qc/0509101}},
	[Erratum: Phys.Rev.D 88, 109905 (2013), Erratum: Phys.Rev.D 90, 109905 (2014)]
	
	\bibitem{Speri:2023jte}
	L.~Speri, M.~L. Katz, A.~J.~K. Chua, S.~A. Hughes, N.~Warburton, J.~E.
	Thompson, C.~E.~A. Chapman-Bird  and J.~R. Gair, \emph{{Fast and Fourier:
			Extreme Mass Ratio Inspiral Waveforms in the Frequency Domain}},
	\href{http://www.arXiv.org/abs/2307.12585}{{\tt 2307.12585}}
	
	\bibitem{Datta:2024vll}
	S.~Datta, R.~Brito, S.~A. Hughes, T.~Klinger  and P.~Pani, \emph{{Tidal heating
			as a discriminator for horizons in equatorial eccentric extreme mass ratio
			inspirals}},
	\href{http://www.arXiv.org/abs/2404.04013}{{\tt 2404.04013}}
	
	\bibitem{Goldberger:2004jt}
	W.~D. Goldberger and I.~Z. Rothstein, \emph{{An Effective field theory of
			gravity for extended objects}}, Phys. Rev. D {\bf 73} (2006) 104029,
	\href{http://www.arXiv.org/abs/hep-th/0409156}{{\tt hep-th/0409156}}
	
	\bibitem{Foffa:2011ub}
	S.~Foffa and R.~Sturani, \emph{{Effective field theory calculation of
			conservative binary dynamics at third post-Newtonian order}}, Phys. Rev. D
	{\bf 84} (2011) 044031,
	\href{http://www.arXiv.org/abs/1104.1122}{{\tt 1104.1122}}
	
	\bibitem{Porto:2016pyg}
	R.~A. Porto, \emph{{The effective field theorist\textquoteright{}s approach to
			gravitational dynamics}}, Phys. Rept. {\bf 633} (2016) 1--104,
	\href{http://www.arXiv.org/abs/1601.04914}{{\tt 1601.04914}}
	
	\bibitem{Bern:2019nnu}
	Z.~Bern, C.~Cheung, R.~Roiban, C.-H. Shen, M.~P. Solon  and M.~Zeng,
	\emph{{Scattering Amplitudes and the Conservative Hamiltonian for Binary
			Systems at Third Post-Minkowskian Order}}, Phys. Rev. Lett. {\bf 122} (2019),
	no.~20, 201603,
	\href{http://www.arXiv.org/abs/1901.04424}{{\tt 1901.04424}}
	
	\bibitem{Foffa:2019yfl}
	S.~Foffa, R.~A. Porto, I.~Rothstein  and R.~Sturani, \emph{{Conservative
			dynamics of binary systems to fourth Post-Newtonian order in the EFT approach
			II: Renormalized Lagrangian}}, Phys. Rev. D {\bf 100} (2019), no.~2, 024048,
	\href{http://www.arXiv.org/abs/1903.05118}{{\tt 1903.05118}}
	
	\bibitem{Bern:2019crd}
	Z.~Bern, C.~Cheung, R.~Roiban, C.-H. Shen, M.~P. Solon  and M.~Zeng,
	\emph{{Black Hole Binary Dynamics from the Double Copy and Effective
			Theory}}, JHEP {\bf 10} (2019) 206,
	\href{http://www.arXiv.org/abs/1908.01493}{{\tt 1908.01493}}
	
	\bibitem{Bern:2020buy}
	Z.~Bern, A.~Luna, R.~Roiban, C.-H. Shen  and M.~Zeng, \emph{{Spinning black
			hole binary dynamics, scattering amplitudes, and effective field theory}},
	Phys. Rev. D {\bf 104} (2021), no.~6, 065014,
	\href{http://www.arXiv.org/abs/2005.03071}{{\tt 2005.03071}}
	
	\bibitem{Kalin:2020mvi}
	G.~K\"alin and R.~A. Porto, \emph{{Post-Minkowskian Effective Field Theory for
			Conservative Binary Dynamics}}, JHEP {\bf 11} (2020) 106,
	\href{http://www.arXiv.org/abs/2006.01184}{{\tt 2006.01184}}
	
	\bibitem{Kalin:2020fhe}
	G.~K\"alin, Z.~Liu  and R.~A. Porto, \emph{{Conservative Dynamics of Binary
			Systems to Third Post-Minkowskian Order from the Effective Field Theory
			Approach}}, Phys. Rev. Lett. {\bf 125} (2020), no.~26, 261103,
	\href{http://www.arXiv.org/abs/2007.04977}{{\tt 2007.04977}}
	
	\bibitem{Bern:2020uwk}
	Z.~Bern, J.~Parra-Martinez, R.~Roiban, E.~Sawyer  and C.-H. Shen,
	\emph{{Leading Nonlinear Tidal Effects and Scattering Amplitudes}}, JHEP {\bf
		05} (2021) 188,
	\href{http://www.arXiv.org/abs/2010.08559}{{\tt 2010.08559}}
	
	\bibitem{Goldberger:2020fot}
	W.~D. Goldberger, J.~Li  and I.~Z. Rothstein, \emph{{Non-conservative effects
			on spinning black holes from world-line effective field theory}}, JHEP {\bf
		06} (2021) 053,
	\href{http://www.arXiv.org/abs/2012.14869}{{\tt 2012.14869}}
	
	\bibitem{Liu:2021zxr}
	Z.~Liu, R.~A. Porto  and Z.~Yang, \emph{{Spin Effects in the Effective Field
			Theory Approach to Post-Minkowskian Conservative Dynamics}}, JHEP {\bf 06}
	(2021) 012,
	\href{http://www.arXiv.org/abs/2102.10059}{{\tt 2102.10059}}
	
	\bibitem{Dlapa:2021npj}
	C.~Dlapa, G.~K\"alin, Z.~Liu  and R.~A. Porto, \emph{{Dynamics of binary
			systems to fourth Post-Minkowskian order from the effective field theory
			approach}}, Phys. Lett. B {\bf 831} (2022) 137203,
	\href{http://www.arXiv.org/abs/2106.08276}{{\tt 2106.08276}}
	
	\bibitem{Cho:2021arx}
	G.~Cho, G.~K\"alin  and R.~A. Porto, \emph{{From boundary data to bound states.
			Part III. Radiative effects}}, JHEP {\bf 04} (2022) 154,
	\href{http://www.arXiv.org/abs/2112.03976}{{\tt 2112.03976}},
	[Erratum: JHEP 07, 002 (2022)]
	
	\bibitem{Bern:2021yeh}
	Z.~Bern, J.~Parra-Martinez, R.~Roiban, M.~S. Ruf, C.-H. Shen, M.~P. Solon  and
	M.~Zeng, \emph{{Scattering Amplitudes, the Tail Effect, and Conservative
			Binary Dynamics at O(G4)}}, Phys. Rev. Lett. {\bf 128} (2022), no.~16,
	161103,
	\href{http://www.arXiv.org/abs/2112.10750}{{\tt 2112.10750}}
	
	\bibitem{Cho:2022syn}
	G.~Cho, R.~A. Porto  and Z.~Yang, \emph{{Gravitational radiation from
			inspiralling compact objects: Spin effects to the fourth post-Newtonian
			order}}, Phys. Rev. D {\bf 106} (2022), no.~10, L101501,
	\href{http://www.arXiv.org/abs/2201.05138}{{\tt 2201.05138}}
	
	\bibitem{Bern:2022kto}
	Z.~Bern, D.~Kosmopoulos, A.~Luna, R.~Roiban  and F.~Teng, \emph{{Binary
			Dynamics through the Fifth Power of Spin at O(G2)}}, Phys. Rev. Lett. {\bf
		130} (2023), no.~20, 201402,
	\href{http://www.arXiv.org/abs/2203.06202}{{\tt 2203.06202}}
	
	\bibitem{Bern:2022jvn}
	Z.~Bern, J.~Parra-Martinez, R.~Roiban, M.~S. Ruf, C.-H. Shen, M.~P. Solon  and
	M.~Zeng, \emph{{Scattering amplitudes and conservative dynamics at the fourth
			post-Minkowskian order}}, PoS {\bf LL2022} (2022)
	051
	
	\bibitem{Dlapa:2022lmu}
	C.~Dlapa, G.~K\"alin, Z.~Liu, J.~Neef  and R.~A. Porto, \emph{{Radiation
			Reaction and Gravitational Waves at Fourth Post-Minkowskian Order}}, Phys.
	Rev. Lett. {\bf 130} (2023), no.~10, 101401,
	\href{http://www.arXiv.org/abs/2210.05541}{{\tt 2210.05541}}
	
	\bibitem{Dlapa:2023hsl}
	C.~Dlapa, G.~K\"alin, Z.~Liu  and R.~A. Porto, \emph{{Bootstrapping the
			relativistic two-body problem}}, JHEP {\bf 08} (2023) 109,
	\href{http://www.arXiv.org/abs/2304.01275}{{\tt 2304.01275}}
	
	\bibitem{Chia:2024bwc}
	H.~S. Chia, Z.~Zhou  and M.~M. Ivanov, \emph{{Bring the Heat: Tidal Heating
			Constraints for Black Holes and Exotic Compact Objects from the
			LIGO-Virgo-KAGRA Data}},
	\href{http://www.arXiv.org/abs/2404.14641}{{\tt 2404.14641}}
	
	\bibitem{Buonanno:1998gg}
	A.~Buonanno and T.~Damour, \emph{{Effective one-body approach to general
			relativistic two-body dynamics}}, Phys. Rev. D {\bf 59} (1999) 084006,
	\href{http://www.arXiv.org/abs/gr-qc/9811091}{{\tt gr-qc/9811091}}
	
	\bibitem{Damour:2001tu}
	T.~Damour, \emph{{Coalescence of two spinning black holes: an effective
			one-body approach}}, Phys. Rev. D {\bf 64} (2001) 124013,
	\href{http://www.arXiv.org/abs/gr-qc/0103018}{{\tt gr-qc/0103018}}
	
	\bibitem{Buonanno:2007pf}
	A.~Buonanno, Y.~Pan, J.~G. Baker, J.~Centrella, B.~J. Kelly, S.~T. McWilliams
	and J.~R. van Meter, \emph{{Toward faithful templates for non-spinning binary
			black holes using the effective-one-body approach}}, Phys. Rev. D {\bf 76}
	(2007) 104049,
	\href{http://www.arXiv.org/abs/0706.3732}{{\tt 0706.3732}}
	
	\bibitem{Pan:2011gk}
	Y.~Pan, A.~Buonanno, M.~Boyle, L.~T. Buchman, L.~E. Kidder, H.~P. Pfeiffer  and
	M.~A. Scheel, \emph{{Inspiral-merger-ringdown multipolar waveforms of
			nonspinning black-hole binaries using the effective-one-body formalism}},
	Phys. Rev. D {\bf 84} (2011) 124052,
	\href{http://www.arXiv.org/abs/1106.1021}{{\tt 1106.1021}}
	
	\bibitem{Pan:2013rra}
	Y.~Pan, A.~Buonanno, A.~Taracchini, L.~E. Kidder, A.~H. Mrou\'e, H.~P.
	Pfeiffer, M.~A. Scheel  and B.~Szil\'agyi, \emph{{Inspiral-merger-ringdown
			waveforms of spinning, precessing black-hole binaries in the
			effective-one-body formalism}}, Phys. Rev. D {\bf 89} (2014), no.~8, 084006,
	\href{http://www.arXiv.org/abs/1307.6232}{{\tt 1307.6232}}
	
	\bibitem{NIST:DLMF}
	\emph{{\it NIST Digital Library of Mathematical Functions}t.}
	\url{https://dlmf.nist.gov/}, Release 1.1.12 of 2023-12-15.
	\newblock F.~W.~J. Olver, A.~B. {Olde Daalhuis}, D.~W. Lozier, B.~I. Schneider,
	R.~F. Boisvert, C.~W. Clark, B.~R. Miller, B.~V. Saunders, H.~S. Cohl, and
	M.~A. McClain, eds.
	
	\bibitem{Flanagan:2012kg}
	E.~E. Flanagan, S.~A. Hughes  and U.~Ruangsri, \emph{{Resonantly enhanced and
			diminished strong-field gravitational-wave fluxes}}, Phys. Rev. D {\bf 89}
	(2014), no.~8, 084028,
	\href{http://www.arXiv.org/abs/1208.3906}{{\tt 1208.3906}}
	
	\bibitem{Brink:2013nna}
	J.~Brink, M.~Geyer  and T.~Hinderer, \emph{{Orbital resonances around Black
			holes}}, Phys. Rev. Lett. {\bf 114} (2015), no.~8, 081102,
	\href{http://www.arXiv.org/abs/1304.0330}{{\tt 1304.0330}}
	
	\bibitem{Nasipak:2021qfu}
	Z.~Nasipak and C.~R. Evans, \emph{{Resonant self-force effects in
			extreme-mass-ratio binaries: A scalar model}}, Phys. Rev. D {\bf 104} (2021),
	no.~8, 084011,
	\href{http://www.arXiv.org/abs/2105.15188}{{\tt 2105.15188}}
	
	\bibitem{Destounis:2021rko}
	K.~Destounis and K.~D. Kokkotas, \emph{{Gravitational-wave glitches: Resonant
			islands and frequency jumps in nonintegrable extreme-mass-ratio inspirals}},
	Phys. Rev. D {\bf 104} (2021), no.~6, 064023,
	\href{http://www.arXiv.org/abs/2108.02782}{{\tt 2108.02782}}
	
	\bibitem{Speri:2021psr}
	L.~Speri and J.~R. Gair, \emph{{Assessing the impact of transient orbital
			resonances}}, Phys. Rev. D {\bf 103} (2021), no.~12, 124032,
	\href{http://www.arXiv.org/abs/2103.06306}{{\tt 2103.06306}}
	
	\bibitem{Nasipak:2022xjh}
	Z.~Nasipak, \emph{{Adiabatic evolution due to the conservative scalar
			self-force during orbital resonances}}, Phys. Rev. D {\bf 106} (2022), no.~6,
	064042,
	\href{http://www.arXiv.org/abs/2207.02224}{{\tt 2207.02224}}
	
	\bibitem{Pan:2023wau}
	Z.~Pan, H.~Yang, L.~Bernard  and B.~Bonga, \emph{{Resonant dynamics of extreme
			mass-ratio inspirals in a perturbed Kerr spacetime}}, Phys. Rev. D {\bf 108}
	(2023), no.~10, 104026,
	\href{http://www.arXiv.org/abs/2306.06576}{{\tt 2306.06576}}
	
\end{thebibliography}

\end{document}